\documentclass[preprint,3p,12pt]{elsarticle}
\usepackage{mathrsfs}
\usepackage{amsmath}
\usepackage{stmaryrd}
\usepackage{bbding}
\usepackage{dcolumn}
\usepackage{graphicx}
\usepackage{amsfonts}
\usepackage{amssymb}
\usepackage{psfrag}
\usepackage{wrapfig}
\usepackage{subfigure}
\usepackage{makeidx}
\usepackage{bm}
\usepackage{epsf}
\usepackage{epsfig}
\usepackage{setspace}
\usepackage{graphicx}
\usepackage{epstopdf}
\usepackage{psfrag}
\usepackage{subfigure}
\usepackage{color}
\begin{document}
\title{Three dimensional high-order gas-kinetic scheme for supersonic isotropic turbulence II: coarse-grained analysis of compressible $K_{sgs}$ budget}

\author[HKUST1]{Guiyu Cao}
\ead{gcaoaa@connect.ust.hk}

\author[BNU]{Liang Pan}
\ead{panliang@bnu.edu.cn}

\author[HKUST1,HKUST2]{Kun Xu \corref{cor}}
\ead{makxu@ust.hk}

\address[HKUST1]{Department of Mathematics, Hong Kong University of Science and Technology, Clear Water Bay, Kowloon, Hong Kong}
\address[BNU]{School of Mathematical Sciences, Beijing Normal University, Beijing, China}
\address[HKUST2]{Shenzhen Research Institute, Hong Kong University of Science and Technology, Shenzhen, China}
\cortext[cor]{Corresponding author}

\begin{abstract}
The direct numerical simulation (DNS) of compressible isotropic turbulence up to the supersonic regime $Ma_{t} = 1.2$ has been investigated by high-order gas-kinetic scheme (HGKS) [{\it{Computers}} \& {\it{Fluids, 192, 2019}}]. In this study, the coarse-grained analysis of subgrid-scale (SGS) turbulent kinetic energy $K_{sgs}$ budget is fully analyzed for constructing one-equation SGS model in the compressible large eddy simulation (LES). The  DNS on a much higher turbulent Mach number up to $Ma_{t} = 2.0$ has been obtained by HGKS, which confirms the super robustness of HGKS. Then, the exact compressible SGS turbulent kinetic energy $K_{sgs}$ transport equation is derived with density weighted filtering process.
Based on the compressible $K_{sgs}$ transport equation, the coarse-grained processes are implemented on three sets of unresolved grids with the Box filter. The coarse-grained analysis of compressible $K_{sgs}$ budgets shows that all unresolved source terms are dominant terms in current system. Especially, the magnitude of SGS pressure-dilation term is in the order of SGS solenoidal dissipation term within the initial acoustic time scale. Therefore, it can be concluded that the SGS pressure-dilation term cannot be neglected as the previous work. The delicate coarse-grained analysis of SGS diffusion terms in compressible $K_{sgs}$ equation confirms that both the fluctuation velocity triple correlation term and the pressure-velocity correlation term are dominant terms. Current coarse-grained analysis gives an indication of the order of magnitude of all SGS terms in compressible $K_{sgs}$ budget, which provides a solid basis for compressible LES modeling in high Mach number turbulent flow.
\end{abstract}


\maketitle

\section{Introduction}
The supersonic turbulence plays a key role in a wide range of natural phenomena and engineering applications,
such as interstellar turbulence, hypersonic spacecraft reentry, and nuclear fusion power reactors \cite{aluie2011compressible,kritsuk2013energy}.
Compared with incompressible turbulence, highly compressible turbulent flows are
more complex due to nonlinear coupling of the velocity, density and pressure fields \cite{lele1994compressibility}.
To elucidate the effects of compressibility in the compressible turbulence, the compressible isotropic turbulence is regarded as one of cornerstones \cite{hanifi2012transition,sagaut2008homogeneous,garnier2009large}.
However, for the compressible isotropic turbulence in supersonic regime ($Ma_t \ge 0.8$), the stronger random shocklets and higher spatial-temporal
gradients pose greater difficulties for both  theoretical analyses and  numerical studies than the flow in other regime \cite{sagaut2008homogeneous,kumar2013weno,cao2019three}.
Currently, the study of supersonic regime is much less known and reported, and very few numerical experiments are available \cite{wang2010hybrid,cao2019three,wang2012scaling}.

For compressible turbulence modeling, the large eddy simulation (LES) for high Mach number turbulent flows is also reported rarely. One-equation subgrid-scale (SGS) models have been extensively used in incompressible LES \cite{schumann1975subgrid,yoshizawa1985statistically,krajnovic2002mixed,de2008localized}.
Since the incorporation of history and non-local effects through transport equation related to the residual motions, the one-equation SGS models have shown better performance in the prediction of turbulent flow.
Meanwhile, compared with the abundant research on compressibility correction for the turbulent kinetic energy equation in Reynolds averaged Navier-Stokes (RANS) simulation \cite{sarkar1991analysis,zeman1991decay,sarkar1992pressure,wilcox1992dilatation,el1993second,ristorcelli1997pseudo,wilcox1998turbulence}, for compressible LES, there only exists limited number of research work on compressible one-equation SGS models \cite{yoshizawa1986statistical,pomraning2002dynamic,park2007numerical,chai2012dynamic}.
With the rapid increasing of computational power, it is well known that the LES gradually becomes the workhorse for high-fidelity turbulence simulation from the smooth turbulent flow to the supersonic one \cite{slotnick2014cfd}.
However, as far as we know, the compressible LES models are less reported, where the algebraic eddy viscosity model can be hardly incorporated with the compressible effect systematically \cite{smagorinsky1963general,germano1991dynamic}.
In the modeling of the compressible effect, it is natural to extend the one-equation SGS model to high turbulent Mach number flow. For compressible one-equation SGS model, an important issue that has not been resolved in the earlier studies is how to distinguish the dominant terms and negligible ones.
Very few coarse-grained analysis of compressible turbulence has been carried out in LES \cite{vreman1995priori,vreman1995direct,martin2000subgrid}, where most of them are limited to the subsonic turbulent Mach number ($Ma_t \le 0.8$).
The priori tests using direct numerical simulation (DNS) data for the calculation of a mixing layer up to Mach number $0.6$ \cite{vreman1995priori,vreman1995direct}, and the DNS for the homogeneous isotropic turbulence up to $Ma_t = 0.52$, were filtered, and the unclosed terms in the momentum, internal energy, and total energy equations were computed \cite{martin2000subgrid}.
It is emphasized that the unresolved dilational dissipation rate and the unresolved pressure-dilation term are important to the compressible LES.
For the forced supersonic isotropic turbulence ($Ma_t \approx1.0$), the filtered result of turbulent kinetic energy transfer on unresolved grids has been well studied \cite{wang2018kinetic}.
While, with the orientation of constructing one-equation SGS model for a much higher turbulent Mach number flow, i.e., $Ma_t \ge 1.0$, the detailed analysis of coarse-grained turbulent kinetic energy budget $K_{sgs}$ is much required in the modeling.

In the past decades, the gas-kinetic scheme (GKS) based on the Bhatnagar-Gross-Krook (BGK) model \cite{bhatnagar1954model,chapman1970mathematical} has been developed systematically for the computations from low speed flow to hypersonic one \cite{xu2001gas,xu2015direct,cao2018physical}.
With the multi-stage multi-derivative framework \cite{li2016two}, a reliable GKS has been constructed with fourth-order and even higher-order accuracy with the implementation of the traditional second-order or third-order flux functions
\cite{pan2016efficient,pan2018two,ji2018family,zhao2019compact}.
In recent years, GKS has been applied in high-Reynolds number turbulent flow \cite{tan2018gas,cao2019implicit}.
More importantly, considering the high-order accuracy in the coupled evolution in space and time, and the super robustness of high-order gas-kinetic scheme (HGKS), the HGKS has been used in the DNS for compressible isotropic turbulence up to the supersonic regime $Ma_{t} = 1.2$ \cite{cao2019three}.  This study confirms that HGKS provides a valid tool for supersonic isotropic turbulence simulation,
and the criterion for a correct DNS solution is determined.
Following the first part \cite{cao2019three}, in order to construct one-equation SGS model for compressible LES, the coarse-grained analysis on supersonic isotropic turbulence is studied here.
In this paper, the DNS on a much higher turbulent Mach number ($Ma_{t} = 2.0$) has been conducted, which confirms the super robustness of HGKS.
Then, the exact compressible turbulent kinetic energy $K_{sgs}$ transport equation has been derived through a density weighted filtering process.
Based on the high-fidelity DNS data, coarse-graining processes are implemented in physical space with a Box filter.
The coarse-grained compressible $K_{sgs}$ budget is fully analyzed and the dominant terms are categorized.
Current coarse-grained analysis provides a solid basis for the compressible LES modeling in the high Mach number turbulent flow.

This paper is organized as follows.
In Section 2, the DNS of supersonic isotropic turbulence by HGKS will be presented.
Section 3 presents the transport equation for the compressible SGS turbulent kinetic energy $K_{sgs}$,
and the implementation of coarse-grained analysis on unresolved grids.
Conclusion is drawn in the final section.

\section{DNS of supersonic isotropic turbulence}
The decaying compressible isotropic turbulence is regarded as one of fundamental benchmarks to study the compressible effect \cite{lele1994compressibility,sagaut2008homogeneous,samtaney2001direct}.
The flow domain of numerical simulation is a cube box defined as $[-\pi, \pi]
\times [-\pi, \pi] \times [-\pi, \pi]$, with periodic boundary
conditions in all three Cartesian directions for all the flow
variables. Evolution of this artificial system is determined by
initial thermodynamic quantities and two dimensionless parameters,
i.e. the initial Taylor microscale Reynolds number
$Re_{\lambda}=\left\langle \rho \right\rangle U_{rms}\lambda/\left\langle \mu \right\rangle$
and turbulent Mach number
$Ma_t=\sqrt{3} U_{rms}/\left\langle  c_s \right\rangle$,
where $\left\langle \cdot \right\rangle$ is the ensemble over the
whole computational domain, $\rho$ is the density, $\lambda$ is the Taylor microscale, $\mu$ is the initial dynamic viscosity, $c_s$ is the sound speed and $U_{rms}$ is
the root mean square of initial turbulent velocity component
$U_{rms}=\left\langle \boldsymbol{U} \cdot \boldsymbol{U}/{3}\right\rangle^{1/2}.$
A three-dimensional solenoidal random initial velocity field
$\boldsymbol{U}$ can be generated by a specified spectrum \cite{passot1987numerical}, which is given by
\begin{align}\label{initial_spectrum}
E(\kappa) = A_0 \kappa^4 \exp (-2\kappa^2/\kappa_0^2),
\end{align}
where $A_0$ is a constant to get a specified initial kinetic energy,
$\kappa$ is the wave number, $\kappa_0$ is the wave number at which
the spectrum peaks. In this paper, fixed $A_0$ and $\kappa_0$  in
Eq.(\ref{initial_spectrum}) are chosen for all cases, which are
initialized by $A_0 = 0.00013$ and $ \kappa_0 = 8$.

Initial strategies play an important role in compressible isotropic
turbulence simulation \cite{samtaney2001direct}, especially for the
starting fast transient period during which the divergence of the
velocity increases rapidly and the negative temperature or pressure
often appear. In the computation, the initial pressure $p_0$, density $\rho_0$ and
temperature $T_0$ are set as constant. In this way, the initial Taylor
microscale Reynolds number $Re_{\lambda}$ and turbulent Mach number
$Ma_{t}$ can be determined by
\begin{align*}
Re_{\lambda}=&\frac{(2 \pi)^{1/4}}{4} \frac{\rho_0}{\mu_0}\sqrt{2
	A_0}\kappa_0^{3/2},\\
Ma_{t}&=\frac{\sqrt{3}}{\sqrt{\gamma R T_0}} U_{rms},
\end{align*}
where the initial density $\rho_0 = 1$, $\mu_0, T_0$ can be
determined by $Re_{\lambda}$ and $Ma_{t}$, and $\gamma = 1.4$ is the
specific heat ratio. In the simulation, the dynamic velocity is
given by $\mu = \mu_0 \big({T}/{T_0}\big)^{0.76}.$
With current initial strategy, the initial ensemble turbulent
kinetic energy $K_0$, ensemble enstrophy $\Omega_0$,  large-eddy-turnover time
$\tau_{to}$, ensemble dissipation rate $\varepsilon_0$, Kolmogorov length scale $\eta_0$, and the Kolmogorov
time scale $\tau_0$ are given as
\begin{equation}\label{initial_def}
\begin{aligned}
K_0=&\frac{3A_0}{64} \sqrt{2 \pi} \kappa_0^5, ~ \Omega_0=\frac{15
	A_0}{256} \sqrt{2 \pi} \kappa_0^7,~
\tau_{to}=\sqrt{\frac{32}{A_0}}(2 \pi)^{1/4} \kappa_0^{-7/2},\\
&\varepsilon_0=2\frac{\mu_0}{\rho_0} \Omega_0, ~
\eta_0=(\nu_0^3/\varepsilon_0)^{1/4}, ~
\tau_0=(\nu_0/\varepsilon_0)^{1/2}.
\end{aligned}
\end{equation}
For decaying compressible isotropic turbulence, the root-mean-square pressure fluctuations $p_{rms}$,
and turbulent kinetic energy $K$ are defined as
\begin{equation}\label{rho_k_prms}
\begin{aligned}
p_{rms}&=\sqrt{\left\langle p -  \left\langle  p \right\rangle \right\rangle}, \\
K&=\frac{1}{2}\left\langle  \rho \boldsymbol{U} \cdot \boldsymbol{U} \right\rangle.
\end{aligned}
\end{equation}
The evolution of turbulent kinetic energy is of interest since it is a fundamental
benchmark for incompressible and compressible turbulence modeling
\cite{lele1994compressibility,yoshizawa1985statistically,pope2001turbulent}. In this study, the ensemble budget of turbulent kinetic energy is computed and analyzed, as the decay of the
ensemble turbulent kinetic energy can be described approximately by
\cite{sarkar1991analysis}
\begin{equation}\label{dkdt}
\begin{aligned}
\frac{\text{d}\left\langle K\right\rangle}{\text{d}t}=\varepsilon&+\left\langle p \theta \right\rangle,\\
\varepsilon=\varepsilon_s+&\varepsilon_d,
\end{aligned}
\end{equation}
where $\varepsilon_s=\left\langle \mu \omega_i
\omega_i\right\rangle$ is the ensemble solenoidal dissipation rate,
$\displaystyle\varepsilon_d= \left\langle 4\mu
\theta^2 /3\right\rangle$ is the ensemble dilational dissipation rate,
$\left\langle p \theta \right\rangle$ is the ensemble
pressure-dilation transfer, $\displaystyle\omega_i=\epsilon_{ijk}
\partial U_k/\partial x_j$ is the fluctuating vorticity,
$\epsilon_{ijk}$ is the alternating tensor, and $\theta = \nabla
\cdot \bm{U}$ is the fluctuating divergence of velocity.

\begin{table}[!h]
	\begin{center}
		\caption{\label{re72gridtable} Parameters for supersonic isotropic turbulence of $R_1$ and $R_2$.}
		\vspace{3mm}
		\centering
		\begin{tabular}{c|cccccc}
			\hline \hline
			Test      &\text{grid size}    &$Ma_{t}$  &$Re_{\lambda}$   &$\kappa_{max} \eta_0$   &$\text{d}t_{ini}/\tau_{to}$   \\
			\hline
			$R_1$     &$384^3$             &2.0       &72               &2.71             &2.00/1000\\
			\hline
			$R_2$     &$512^3$             &2.0       &120              &2.80             &3.40/1000\\
			\hline \hline
		\end{tabular}
	\end{center}
\end{table}
\begin{figure}[!h]
	\centering
	\includegraphics[width=0.45\textwidth]{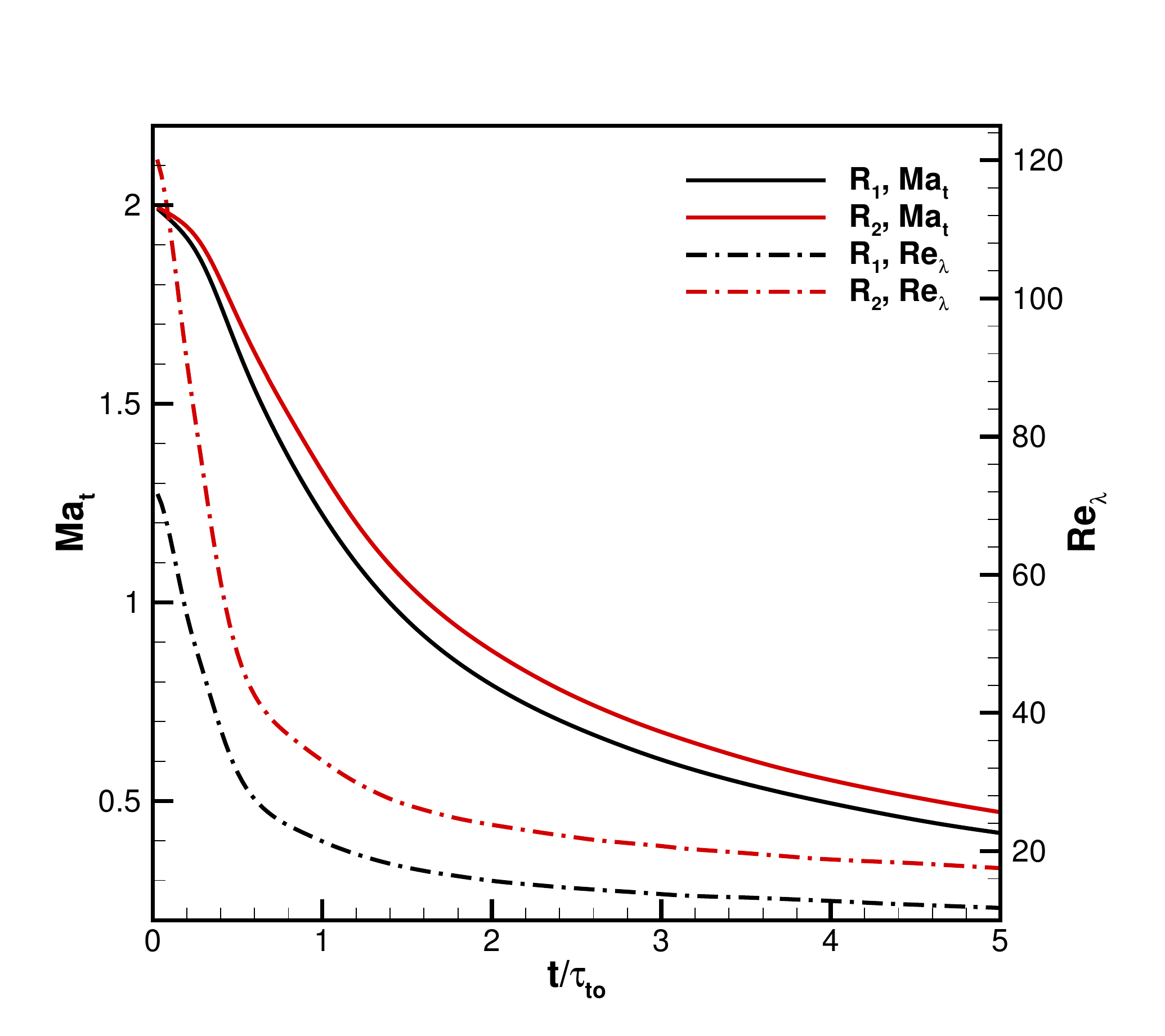}
	\includegraphics[width=0.45\textwidth]{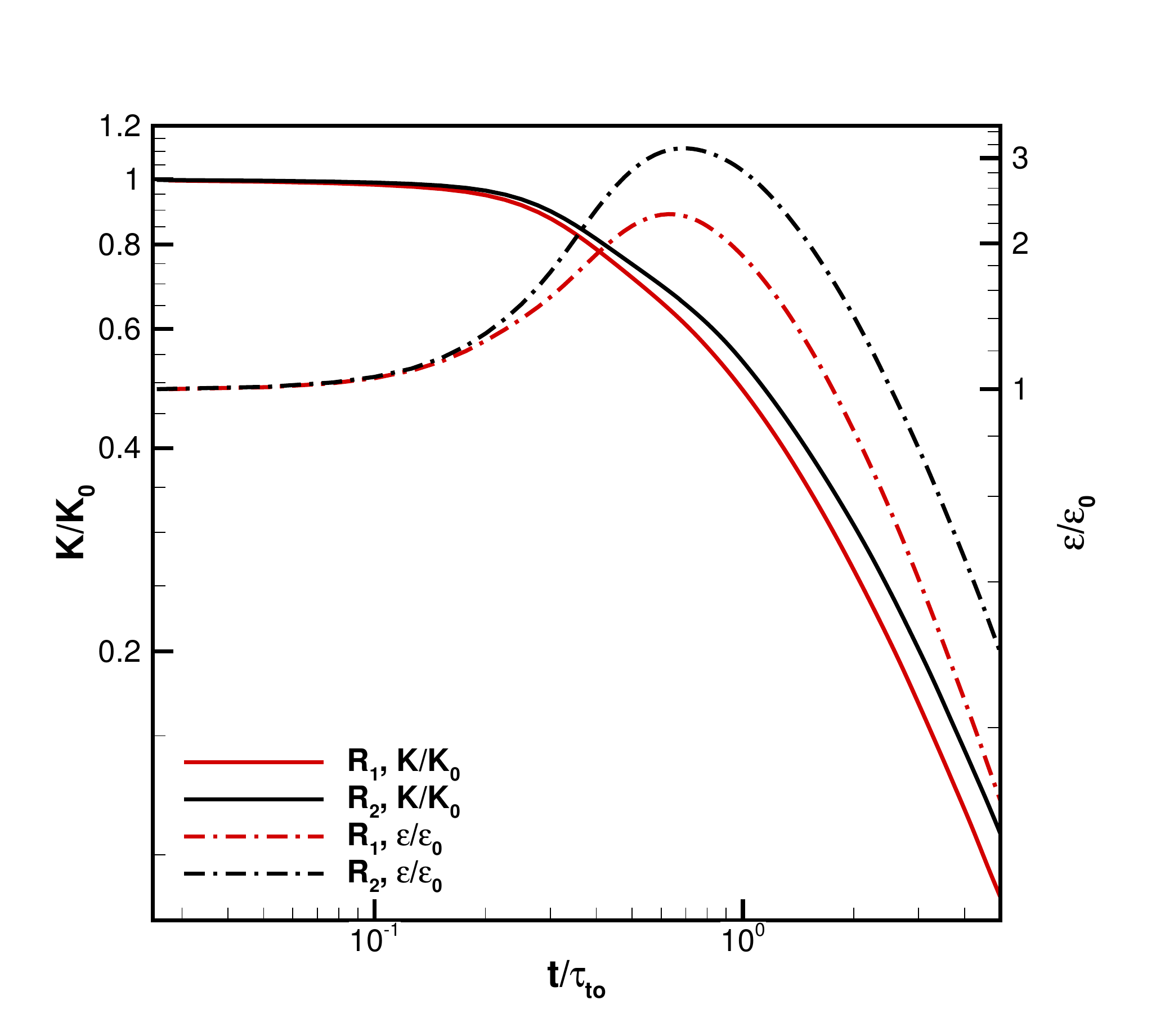}\\
	\includegraphics[width=0.45\textwidth]{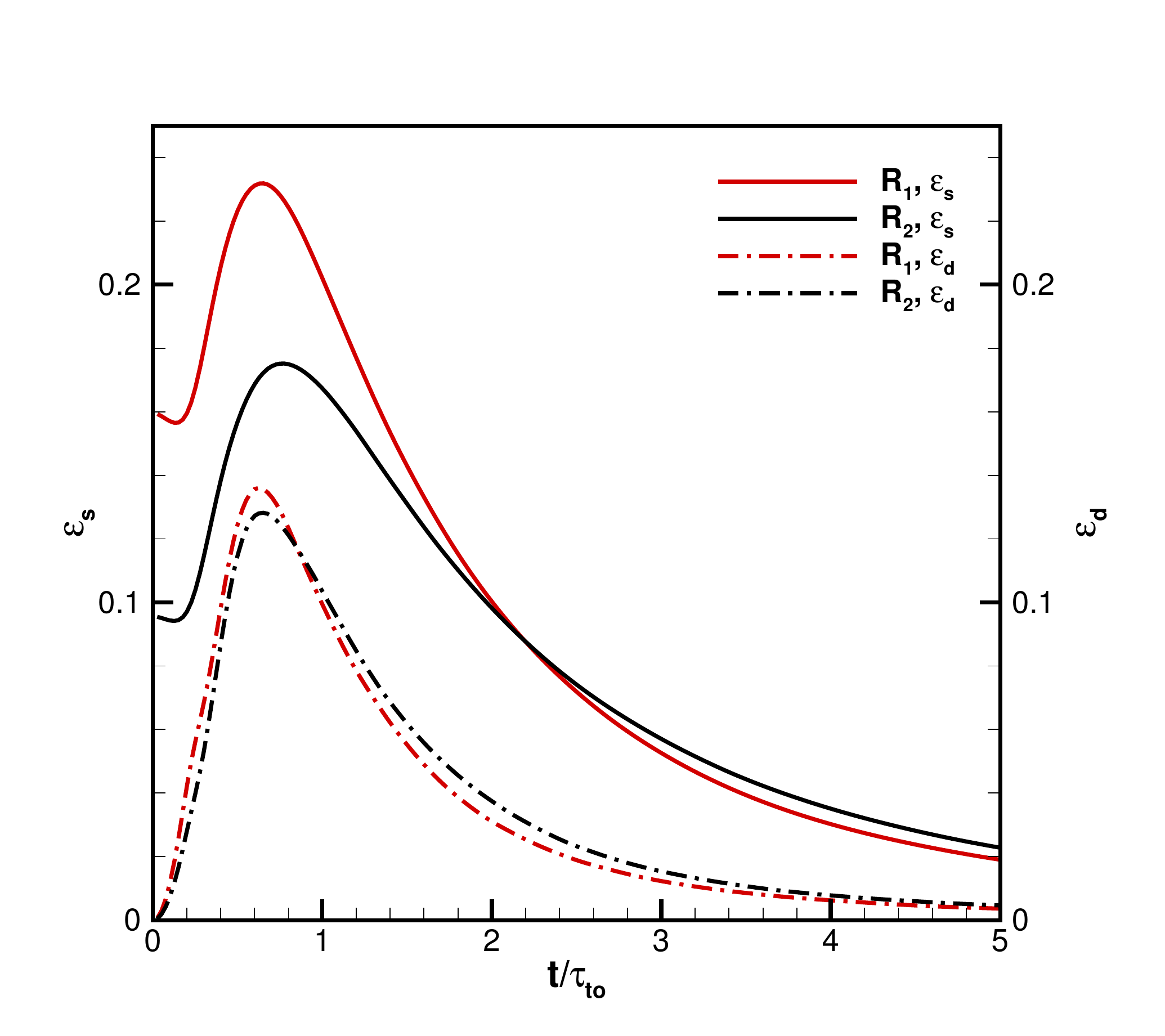}
	\includegraphics[width=0.45\textwidth]{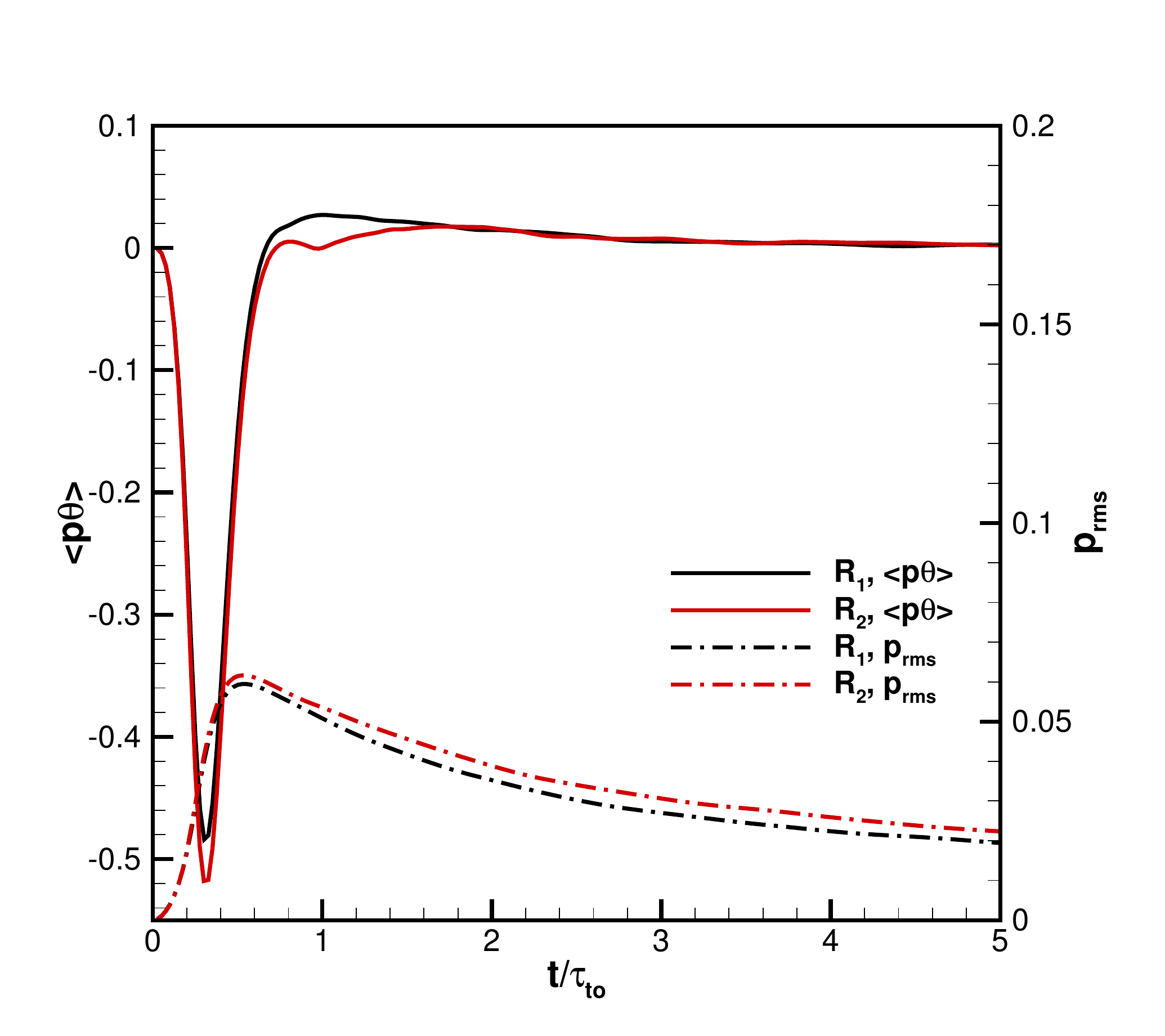}
	\vspace{-2mm}
	\caption{\label{history_R1R2} Time history of $M_t$ and $Re_{\lambda}$,
		$K/K_0$ and $\varepsilon/\varepsilon_0$, $\varepsilon_s$ and
		$\varepsilon_d$, and $\left\langle p \theta \right\rangle$ and $p_{rms}$ for cases
		$R_1$ and $R_2$.}
\end{figure}
In this section, the DNS study of decaying supersonic isotropic turbulence at a fixed turbulent Mach number $Ma_t=2.0$ with Taylor microscale Reynolds number $Re_{\lambda} = 72$ and $Re_{\lambda} = 120$ are implemented.
The grid size and time step are guided by previous criterion of HGKS \cite{cao2019three}.
The details of numerical tests $R_1$ and $R_2$ are given in Table.\ref{re72gridtable}, where $\Delta$ is the uniform grid size in each
direction, $\kappa_{max} = \sqrt{2} \kappa_0 N/3$ is the maximum resolved number wave number \cite{eswaran1988examination}, $\kappa_0 = 8$ in Eq.(\ref{initial_spectrum}) and $N$ is the number of grid points in
each Cartesian direction.
Here $\text{d}t_{ini}$ represents the time step for the initial step, and the initial large-eddy turnover
time $\tau_{to}$ can be determined by Eq.\eqref{initial_def}.
The detailed numerical scheme can be found in the first part of HGKS for supersonic isotropic turbulence \cite{cao2019three,pan2016efficient,pan2018two}.

\vspace{-5mm}
\begin{figure}[!h]
	\centering
	\includegraphics[width=0.45\textwidth]{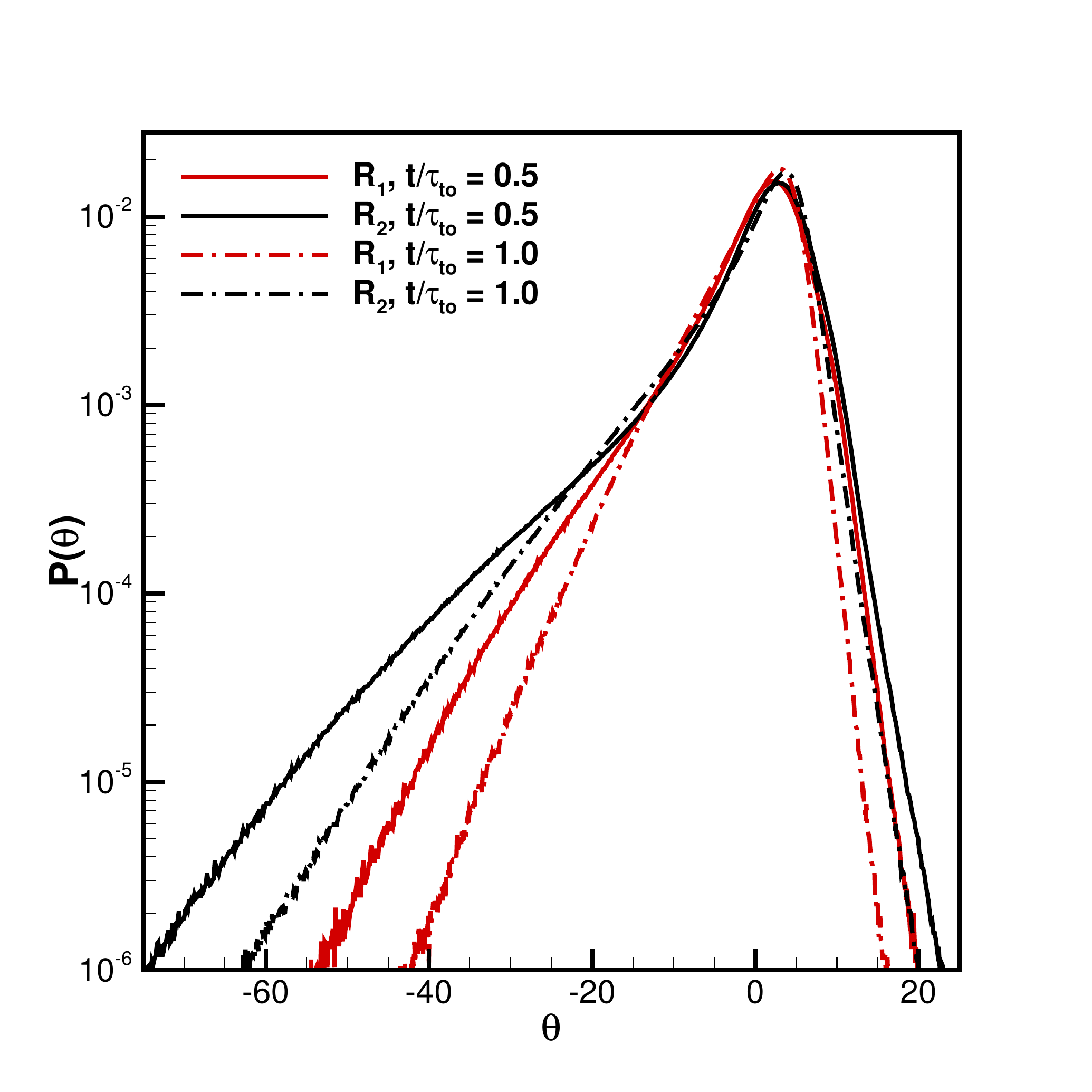}	
	\includegraphics[width=0.45\textwidth]{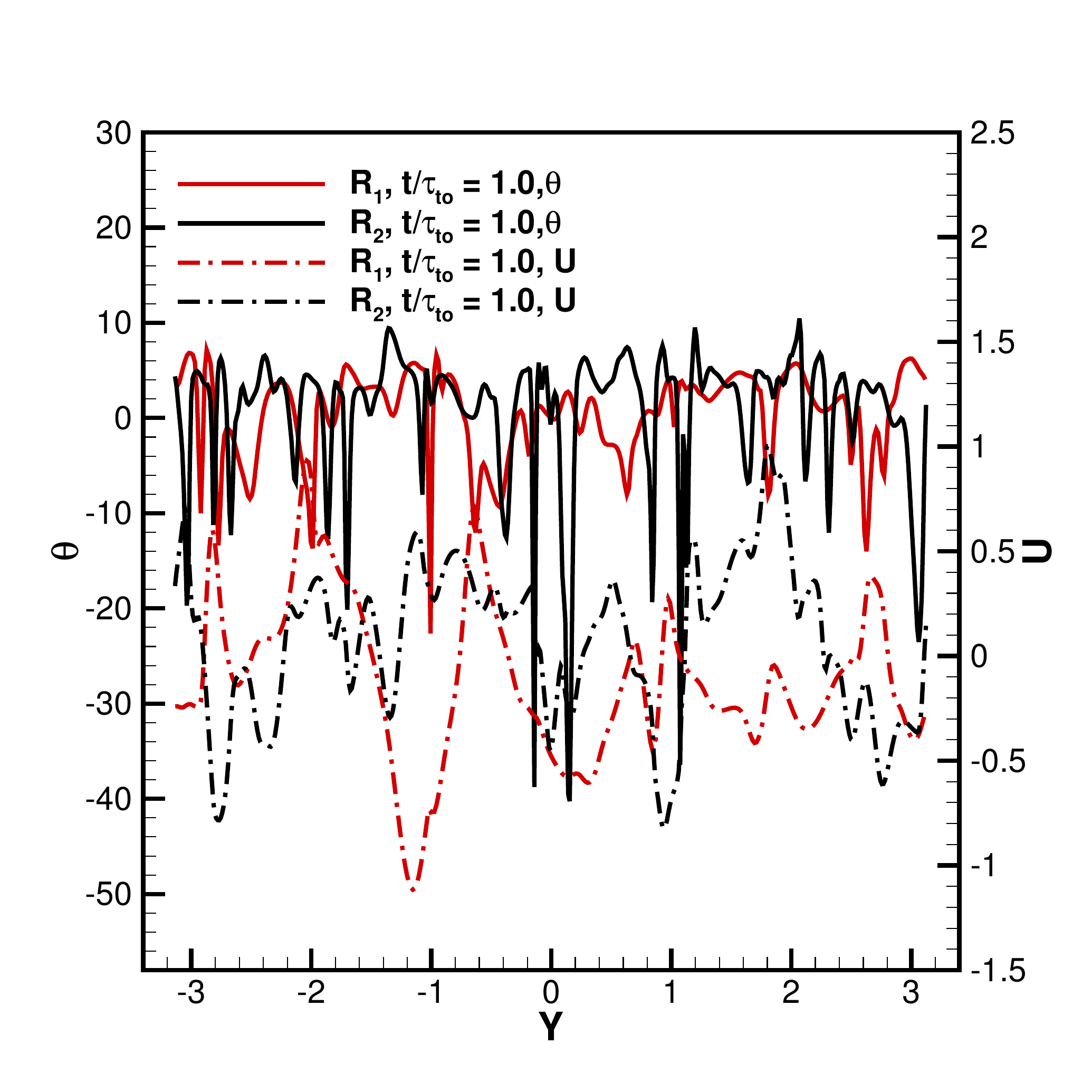}
	\vspace{-3mm}
	\caption{\label{q_pdftheta_R1R2} PDF of dilation $\theta$, $x$-direction velocity component $U$ and dilation $\theta$ along  $x=0$ and $z=0$ at $t/\tau_{to} = 0.5$ and $t/\tau_{to} = 1.0$ for cases $R_1$ and $R_2$.}
	%\vspace{-2mm}
	\includegraphics[width=0.45\textwidth]{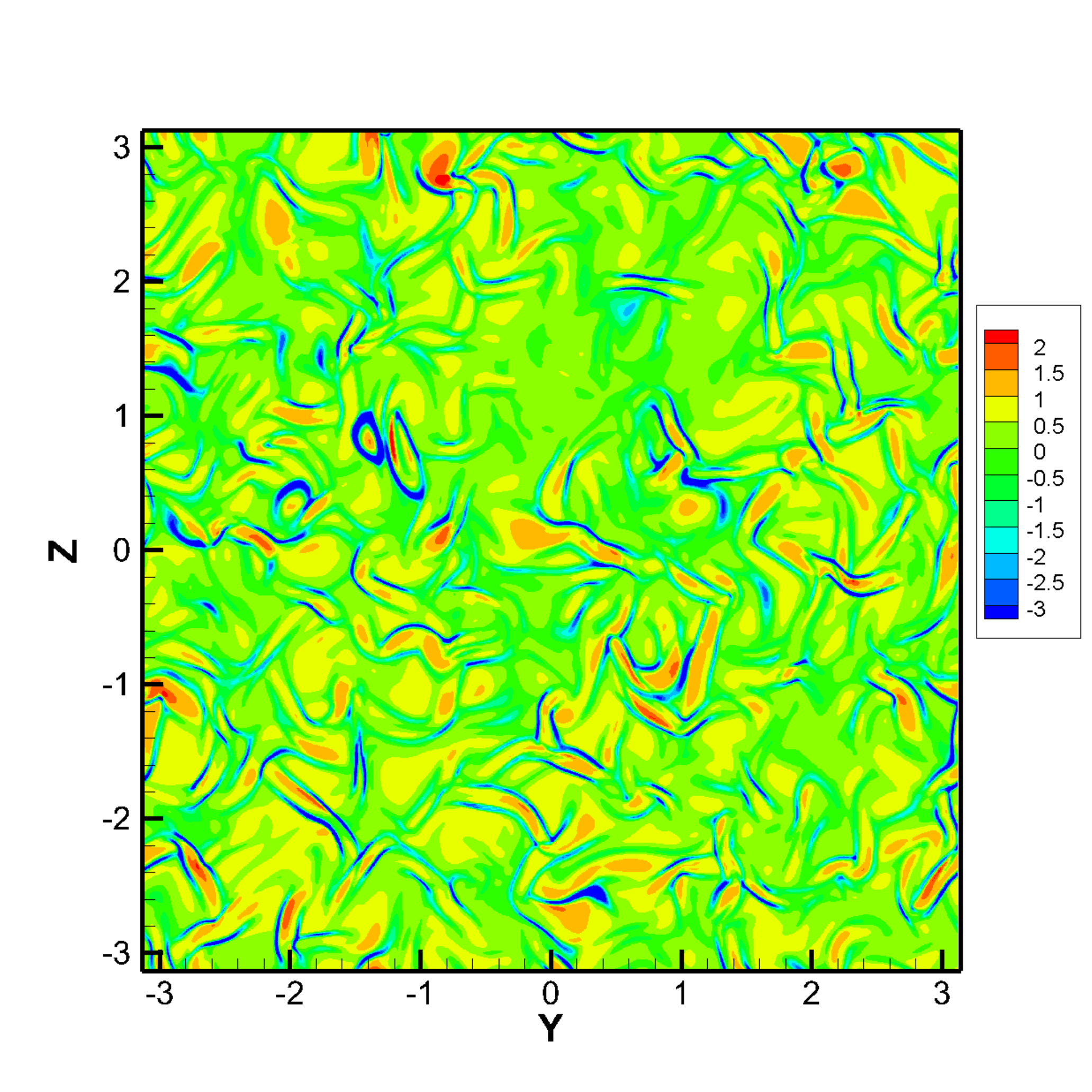}
	\vspace{-3mm}
	\caption{\label{q_pdftheta_R1R2-b} Contour of normalized dilation $\theta/\left\langle \theta
		\right\rangle^{\ast}$ at $x=0$  at $t/\tau_{to}=0.5$ for case $R_1$.}
\end{figure}
\vspace{-2mm}

The time history of statistical quantities in Eq.\eqref{rho_k_prms} and Eq.(\ref{dkdt}) are presented in Fig \ref{history_R1R2}.
The ensemble turbulent Mach number $Ma_t$ and Taylor microscale Reynolds number $Re_{\lambda}$ decay monotonically.
During the early stage, $Re_{\lambda}$ decays very fast.
Up to $t/\tau_{to}=1.0$, the Taylor microscale Reynolds number $Re_{\lambda}$ is approximate $20\%$ of the initial values.
Meanwhile, the ensemble dissipation rate $\varepsilon$ reaches its maximum, which is around $3$ times of $\varepsilon_0$.
Obviously, the peak ensemble dilational dissipation rate $\varepsilon_d$ is approximately half of the peak ensemble solenoidal dissipation rate $\varepsilon_s$, which is the significant behavior of high Mach number turbulent flows.
Additionally, the ensemble dilational dissipation rate depends on $Re_{\lambda}$ slightly, which is confirmed with previous analysis \cite{cao2019three}.
Root-mean-square pressure fluctuations $p_{rms}$ reaches its maximum around $t/\tau_{to}=0.6$, corresponding to the peak ensemble dilational dissipation rate.
During the early stage of the decaying supersonic isotropic turbulence, the ensemble pressure-dilation term can be in the same order of ensemble total dissipation rate \cite{samtaney2001direct}.
The transfer from turbulent kinetic energy to internal energy cannot be neglected as the forced supersonic isotropic turbulence \cite{wang2018kinetic}.
After $t/\tau_{to} \approx 1.0$, $\left\langle p \theta \right\rangle$ changes signs during the evolution and preserves small but positive value.

The probability density functions (PDF) of dilation $\theta$ and $x$-direction velocity component $U$ and dilation $\theta$ are presented in Fig.\ref{q_pdftheta_R1R2}.
All PDFs of dilation $\theta$ in Fig.\ref{q_pdftheta_R1R2} are obtained by dividing the dilation range into $1000$ equivalent intervals.
All PDFs of dilation show strong negative tails, which are the most significant flow structures of compressible isotropic  turbulence resulting from the shocklets \cite{kumar2013weno,cao2019three,wang2010hybrid,wang2012scaling}.
The $x$-direction velocity component $U$ and dilation $\theta$ along the $x= 0$ and $z= 0$ indicates that the strong shocklets and high expansion regions appear frequently and randomly.
Contour of normalized dilation $\theta/\left\langle \theta \right\rangle ^{\ast}$ at $t/\tau_{to} = 0.5$ of $R_1$ is presented in Fig.\ref{q_pdftheta_R1R2-b}.
Contour of normalized dilation shows very different behavior between the compression motion and expansion
motion, where $\left\langle \theta \right\rangle^{\ast}$ is the root-mean-square dilation.
Strong compression regions $\theta/\left\langle \theta \right\rangle^{\ast} \leq -3$ are usually recognized as shocklets \cite{samtaney2001direct}.
These random distributed shocklets and high expansion region lead to strong spatial gradient in flow fields,
which pose much greater challenge for high-order schemes when implementing DNS for isotropic turbulence in supersonic regime.
DNS on a much higher turbulent Mach number up to $Ma_{t} = 2.0$ has been obtained by HGKS, which confirms the super robustness of HGKS.
Based on the high-fidelity DNS data, the coarse-grained analysis for compressible SGS turbulent kinetic energy will be implemented for constructing the compressible one-equation SGS model.

\section{Coarse-grained analysis of compressible $K_{sgs}$ budget}
In this section, the exact compressible SGS turbulent kinetic energy $K_{sgs}$ transport equation will be derived with density weighted filtering process.
The Box filter \cite{piomelli1991subgrid,vreman1994realizability} is used for the coarse-graining processes of compressible $K_{sgs}$ transport equation on three sets of unresolved grids.
Finally, the  dominant terms in compressible $K_{sgs}$ transport equation are determined for constructing the compressible one-equation SGS model for high turbulent Mach number turbulent flows.

\subsection{Compressible $K_{sgs}$ transport equation}
For LES models \cite{smagorinsky1963general,pope2001turbulent}, after filtering process on unresolved grids, the flow variables can be decomposed into resolved (filtered) and SGS (residual) terms as follows
\begin{equation}\label{filtering}
\begin{aligned}
\phi(\boldsymbol{x}) = \overline{\phi} (\boldsymbol{x}) + \phi^{'}(\boldsymbol{x}).
\end{aligned}
\end{equation}
The filtered terms is defined as
\begin{equation*}
\begin{aligned}
\overline{\phi} (\boldsymbol{x}) = \int_{\Omega} G(\boldsymbol{x}, \boldsymbol{x}', \boldsymbol{l}) \phi(\boldsymbol{x}') \text{d}\boldsymbol{x}',
\end{aligned}
\end{equation*}
where  $\Omega$ is the filtered domain and $\boldsymbol{l}$ denotes
the filter width associated with the wavelength of the smallest scale retained by the coarse-graining operation.  The filter function $G$ is defined as
\begin{equation*}
G(\boldsymbol{x}, \boldsymbol{x}', \boldsymbol{l})=\prod_i G_i (x_i, x_i^{\prime},l_i).
\end{equation*}
The following Box filter \cite{pope2001turbulent,piomelli1991subgrid} in physical space is used in this paper
\begin{equation*}\label{box_filter}
\begin{aligned}
G_i (x_i, x_i^{\prime},l_i) =
\begin{cases}
1/l_i, &\text{for}~ |x_i - x_i^{\prime}| \leq l_i/2, \\
0, &\text{otherwise},
\end{cases}
\end{aligned}
\end{equation*}
where $l_i$ is the filter width in $i$-direction, and the positive definite kernel of Box filter
allows positive SGS turbulent kinetic energy \cite{vreman1994realizability}.
Various filter-widths $\l_i = n \Delta_i$ are used in the following analysis,  where $\Delta_i$ is the $i$-direction grid size.
In current study, the filter width and the grid size are equivalent in $x$, $y$ and $z$ directions.
With the filtered process, the one transport equation $K_{sgs}$ of subgrid-scale kinetic energy for incompressible LES \cite{schumann1975subgrid, yoshizawa1985statistically} has been derived.

For compressible turbulence modeling, to avoid subgrid term appearing in the filtered continuity equation, the density-weighted (Favre) filtering \cite{favre1965equations} is applied, which reads
\begin{equation}\label{favre_average}
\begin{aligned}
    \tilde{\phi} = \frac{\overline{ \rho \phi}}{\overline{\rho}}.
\end{aligned}
\end{equation}
In this way, SGS stress $\tau_{ij}$ and SGS kinetic energy $\overline{\rho} K_{sgs}$ are defined as
\begin{equation}\label{les_stress}
\begin{aligned}
    \tau_{ij} &= \overline{\rho} (\widetilde{U_i U_j} - \widetilde{U}_i \widetilde{U}_j), \\
    \overline{\rho} K_{sgs} &= \frac{1}{2} \tau_{kk} = \frac{1}{2} \overline{\rho} (\widetilde{U_k U_k} - \widetilde{U}_k \widetilde{U}_k).
\end{aligned}
\end{equation}
The compressible SGS kinetic energy equation can be derived as Appendix A, the governing equation is given by
\begin{equation}
\begin{aligned}\label{rhok_les}
     (\overline{\rho} K_{sgs})_{,t} + (\overline{\rho} K_{sgs} \widetilde{U}_j)_{,j} \color{black} = P_{sgs} - D_{sgs} + \Pi_{sgs} + T_{sgs},
\end{aligned}
\end{equation}
where $P_{sgs}$ is the SGS production term, $D_{sgs}$ is the SGS dissipation term, $\Pi_{sgs}$ is the SGS pressure dilation term, and the last term $T_{sgs}$ is the sum of SGS diffusion terms.
More specifically, the right-hand-side terms in Eq.(\ref{rhok_les}) can be written as
\begin{equation}
\begin{aligned}\label{rhs_les}
    P_{sgs} &= - \tau_{ij} \widetilde{S}_{ij}, \\
    D_{sgs} &= \overline{\sigma_{ij} U_{i,j}} - \overline{\sigma}_{ij} \widetilde{U}_{i,j}, \\
    \Pi_{sgs} &= \overline{p U_{k,k}} - \overline{p} \widetilde{U}_{k,k}, \\
    T_{sgs} &= [- \frac{1}{2}\overline{\rho} (\widetilde{U_i U_i U_j} - \widetilde{U_i U_i} \tilde{U}_j) + \tau_{ij} \widetilde{U}_i + (\overline{\sigma_{ij} U_i} - \overline{\sigma}_{ij} \widetilde{U}_i) - \overline{\rho} R (\widetilde{T U_j} - \tilde{T} \widetilde{U}_j)]_{,j},
\end{aligned}
\end{equation}
where $\widetilde{S}_{ij} = 1/2(\widetilde{U}_{i,j} + \widetilde{U}_{j,i})$.
More details about the derivation of Eq.\eqref{rhok_les} can be found in Appendix A.
The SGS production term $-\tau_{ij} \widetilde{S}_{ij}$ represents the inter-scale transfer associated with the interaction of the resolved and unresolved scales.
There exists local SGS turbulent kinetic energy backscatter, which illustrates the SGS turbulent kinetic energy transfer from sub-grid scales to resolved scales \cite{wang2018kinetic,piomelli1991subgrid}.
As presented in Appendix A, the total SGS dissipation rate $D_{sgs}$ can be decomposed into two parts, the SGS solenoidal dissipation rate $\varepsilon_s^{sgs}$ and SGS dilational dissipation rate $\varepsilon_d^{sgs}$, as
\begin{equation}
\begin{aligned}\label{dissipation}
    \varepsilon_s^{sgs} &= \overline{\mu} (\widetilde{\omega_i \omega_i} - \tilde{\omega}_i \tilde{\omega}_i), \\
    \varepsilon_d^{sgs} &= 4 \overline{\mu} (\widetilde{U_{k,k}^2} - \widetilde{U}_{k,k}^2)/3,
\end{aligned}
\end{equation}
where $\omega_i =  \epsilon_{ijk} U_{k,j}$ is the resolved vorticity and $\tilde{\omega}_i =  \epsilon_{ijk} \widetilde{U}_{k,j}$ is the unresolved one with the alternating tensor $\epsilon_{ijk}$.
There is a slight difference between Eq.(\ref{dissipation}) and Eq.(3.8) in the reference \cite{chai2012dynamic}.
Restricting the analysis to the linear Kovasznay splitting \cite{kovasznay1953turbulence}, the solenoidal dissipation is associated entirely with the vorticity mode, whereas the dilational dissipation is mainly due to the acoustic mode in the absence of significant entropy source \cite{sagaut2008homogeneous}.
$\Pi_{sgs}$ is SGS pressure-dilation term, which is related to the redistribute $K_{sgs}$ in the flowfields for compressible turbulence.
The SGS pressure-dilation term reduce to $0$ in the incompressible limit.
$T_{sgs}$ is the sum of all SGS diffusion terms, which are usually grouped and modeled together both for incompressible and compressible turbulence models \cite{garnier2009large, wilcox1998turbulence}.
In this paper, to determine the dominant SGS diffusion term, all SGS diffusion terms are analyzed in detail.

According to the Eq.(\ref{rhs_les}), the right-hand-side terms of Eq.(\ref{rhok_les}) are classified as Table.\ref{table_rhs_terms}.  With the Favre filtering process on unresolved grids, the analysis of dominant source terms and SGS diffusion terms will be presented in the following section.
\begin{table}[!h]
	\begin{center}
		\caption{\label{table_rhs_terms} Expressions for the right-hand-side terms in compressible $K_{sgs}$ equation.}
		\vspace{3mm}
		\centering
		\begin{tabular}{cc|cc}
			\hline \hline
			Symbol              &Expression    &Symbol          &Expression     \\
			\hline
			$P$         &$- \tau_{ij} \widetilde{S}_{ij} \color{black}$         &$T_1$       &$[- \frac{1}{2}\overline{\rho} (\widetilde{U_i U_i U_j} - \widetilde{U_i U_i} \tilde{U}_j)]_{,j}$          \\
			$D_1$       &$\overline{\mu} (\widetilde{\omega_i \omega_i} - \tilde{\omega}_i \tilde{\omega}_i)$   &$T_2$       &$(\tau_{ij} \tilde{U}_i)_{,j}$    \\
			$D_2$       &$4 \overline{\mu} (\widetilde{U_{k,k}^2} - \widetilde{U}_{k,k}^2)/3$   &$T_3$       &$[(\overline{\sigma_{ij} U_i} - \overline{\sigma}_{ij} \tilde{U}_i)]_{,j}$    \\
			$\Pi$       &$\overline{p U_{k,k}} - \overline{p} \widetilde{U}_{k,k}$   &$T_4$       &$[- \overline{\rho} R (\widetilde{T U_j} - \tilde{T} \tilde{U}_j)]_{,j}$    \\
			\hline \hline
		\end{tabular}
	\end{center}
	\vspace{-5mm}
\end{table}

\subsection{Coarse-grained analysis of compressible $K_{sgs}$ transport equation}
The DNS and filtering LES grids for $R_1$ and $R_2$ are presented in Table.\ref{caseAB}. The discretization method of spatial derivatives plays a key role in analyzing the budget of
compressible $K_{sgs}$ transport equation. In current paper, to be consistent with HGKS calculation \cite{cao2019three}, the fifth-order WENO-Z reconstruction \cite{castro2011high} is adopted in computing the spatial derivatives of flow variables, and details are given in Appendix B.
\begin{table}[!h]
	\caption{DNS and filtering LES grids for $R_1$ and $R_2$}
	\vspace{3mm}
	\centering
	\begin{tabular}{c|cc|c|cc}
		\hline \hline
		&grid size            & $\kappa_{max}\eta_0$ &                    &grid size & $\kappa_{max}\eta_0$ \\
		\hline
		DNS                   &$384^3$         &$2.71$  &DNS                  &$512^3$         &$2.80$   \\
		\hline
		case $A_1$            &$192^3$         &$1.36$  &case $B_1$           &$256^3$         &$1.40$   \\
		\hline
		case $A_2$            &$96^3$          &$0.68$  &case $B_2$           &$128^3$         &$0.70$   \\
		\hline
		case $A_3$            &$64^3$          &$0.45$  &case $B_3$           &$64^3$          &$0.35$   \\
		\hline \hline
	\end{tabular}
	\label{caseAB}
	\vspace{-1mm}
\end{table}

\begin{figure}[!htp]
	\centering
	\includegraphics[width=0.32\textwidth]{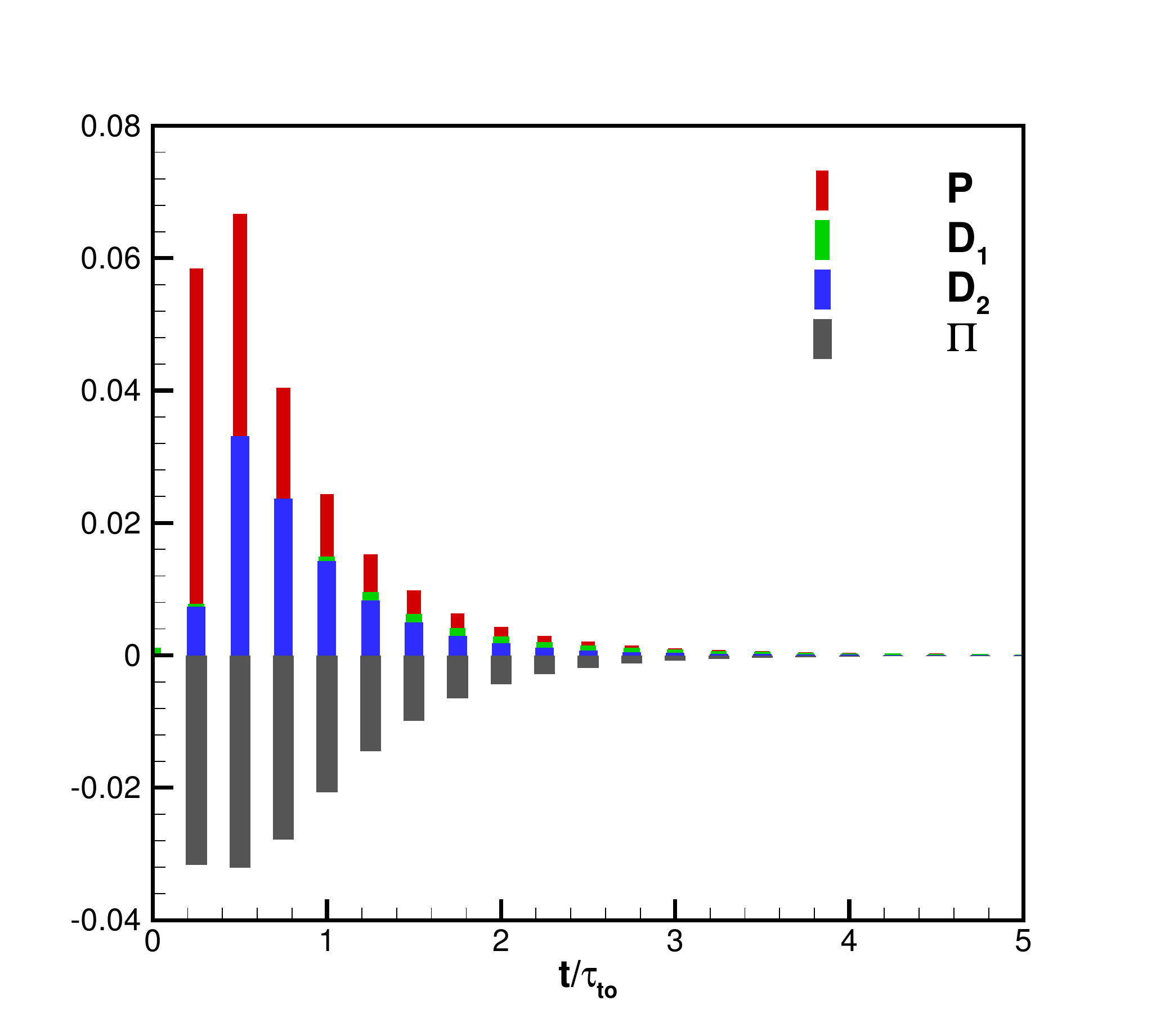}
	\includegraphics[width=0.32\textwidth]{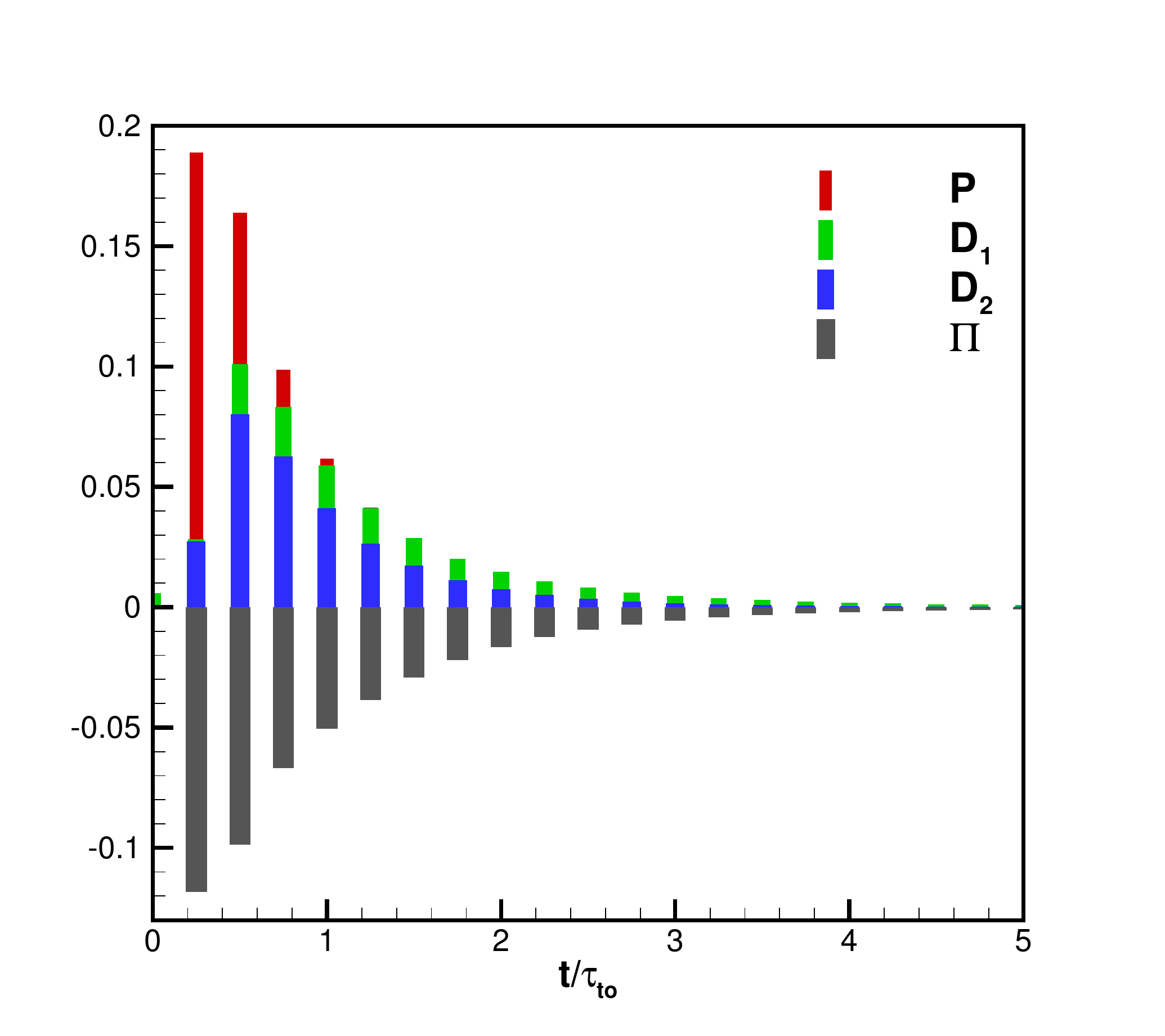}
	\includegraphics[width=0.32\textwidth]{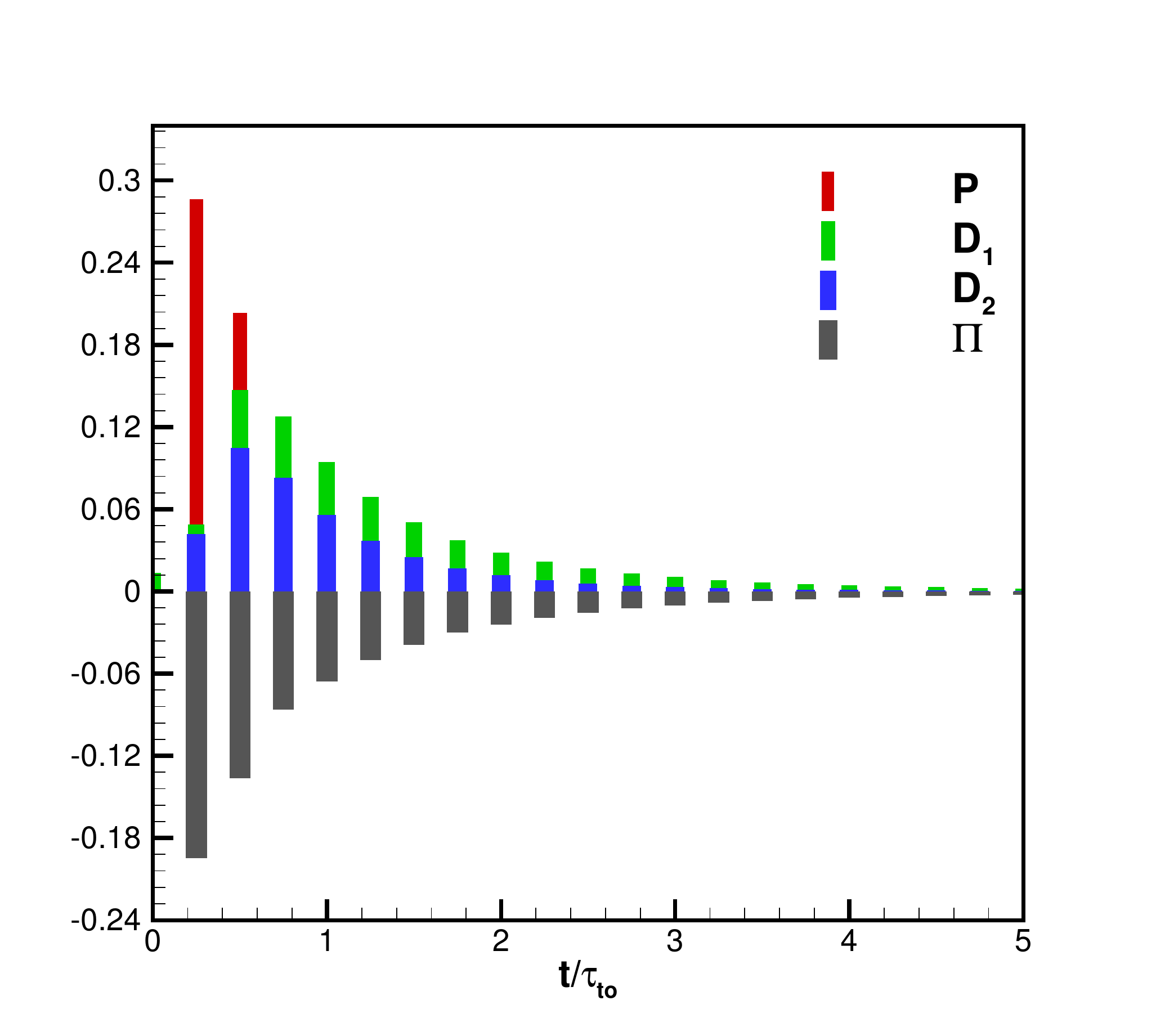}
	\includegraphics[width=0.32\textwidth]{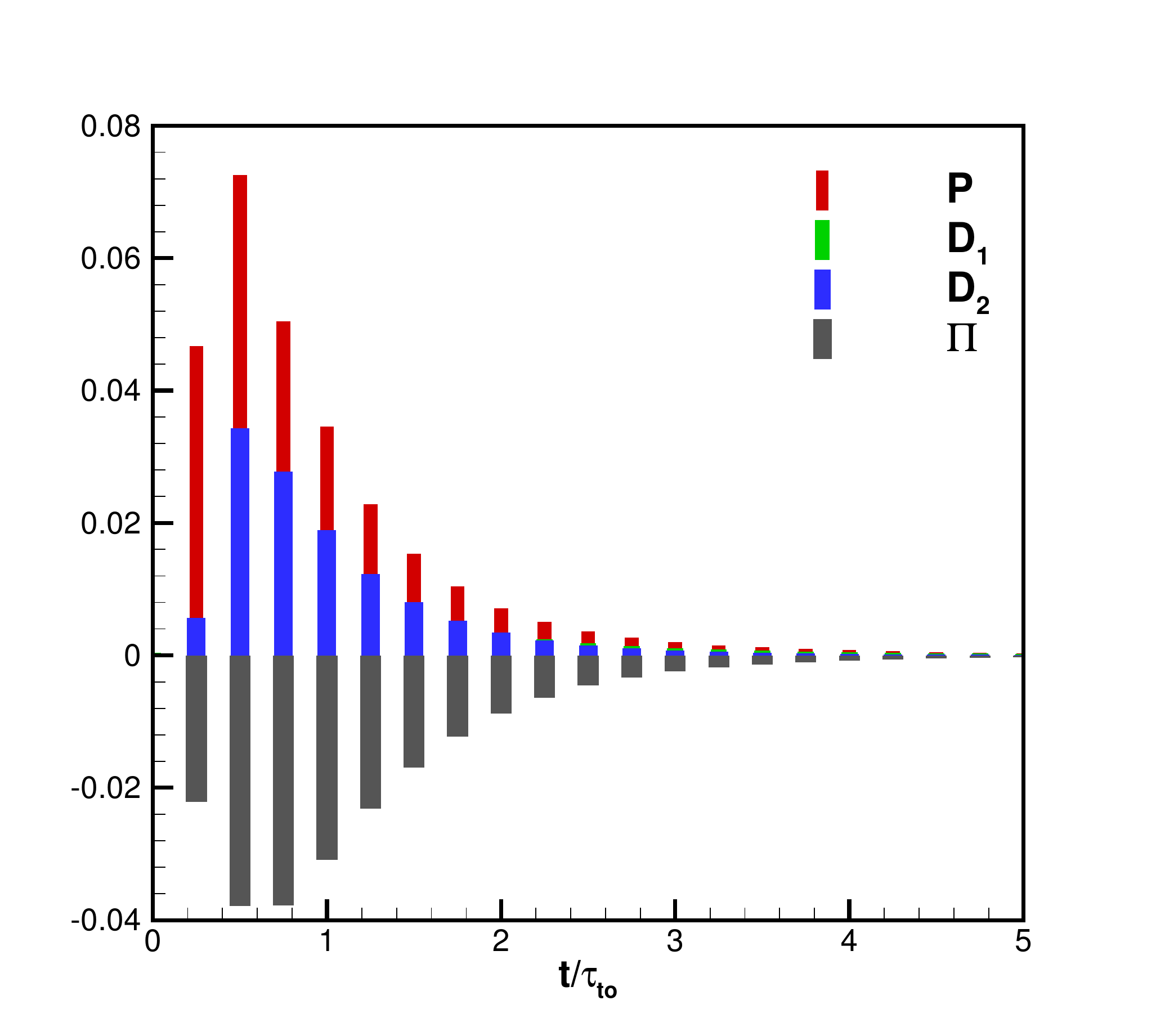}
	\includegraphics[width=0.32\textwidth]{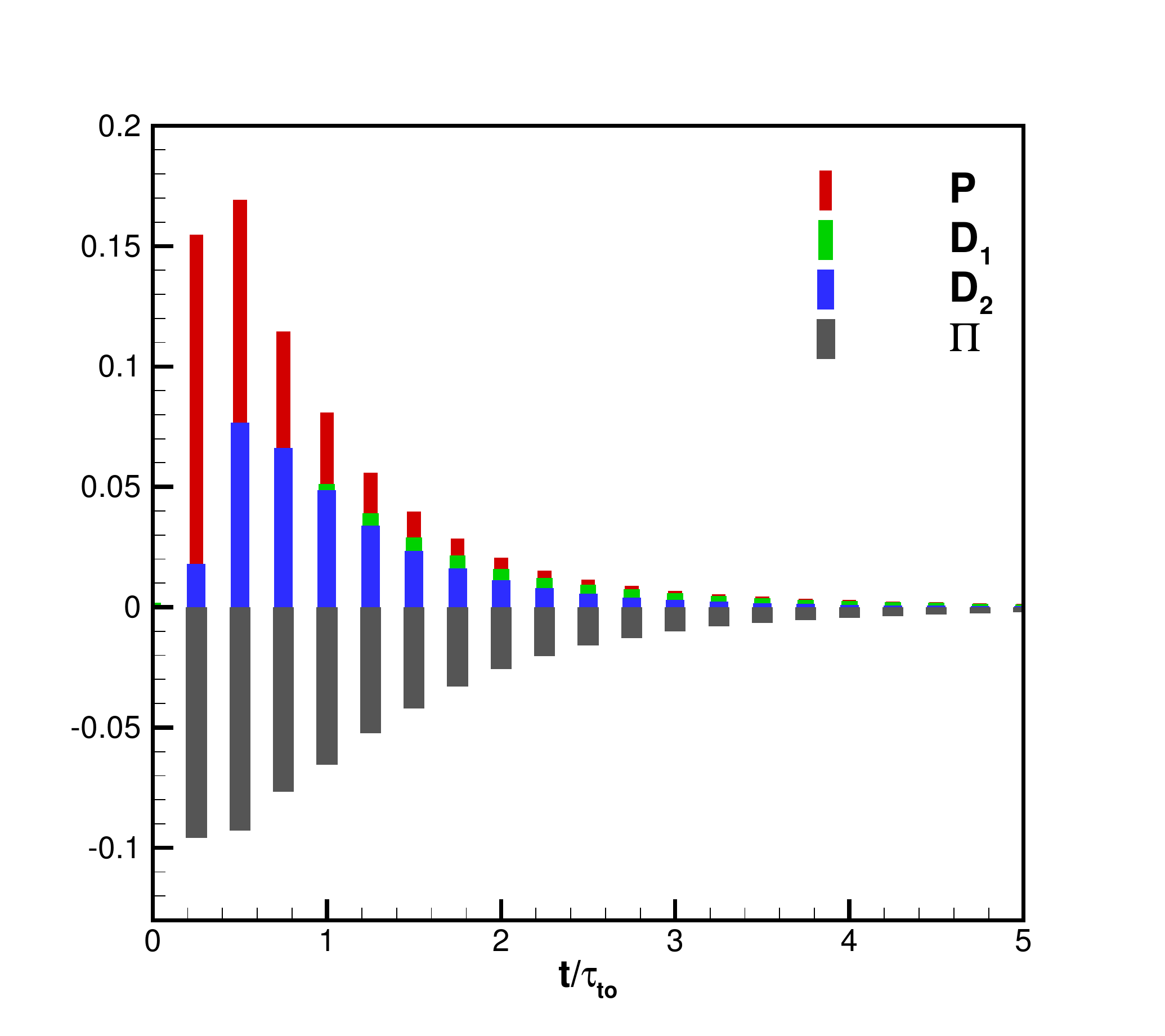}
	\includegraphics[width=0.32\textwidth]{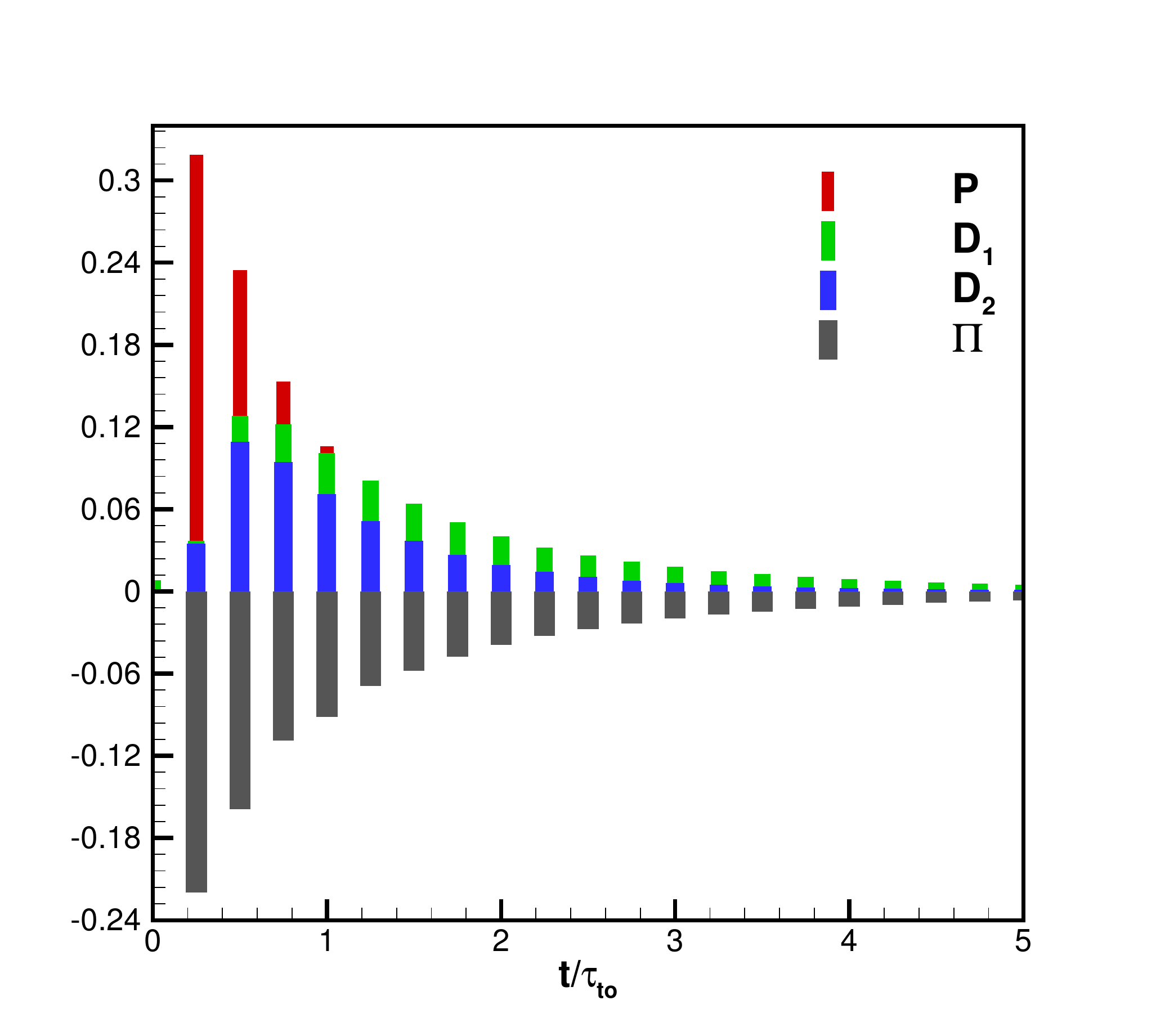}
	\vspace{-2mm}
	\caption{\label{Ksgs_budget_ensemble_sum} Coarse-grained compressible $K_{sgs}$ budgets of $P$, $D_1$, $D_2$, and $\Pi$ for cases $A_1$, $A_2$, $A_3$ (upper row) and $B_1$, $B_2$, $B_3$ (lower row). }
	\vspace{-2mm}
\end{figure}
The coarse-grained compressible $K_{sgs}$ budgets $P$, $D_1$, $D_2$ and $\Pi$  in Eq.(\ref{rhs_les}) for cases $A_1$-$A_3$ and $B_1$-$B_3$ are presented as the Fig.{\ref{Ksgs_budget_ensemble_sum}}.
The budgets are computed in the ensemble norm, and the spatial derivatives are obtained by WENO-Z reconstruction as the Appendix B.
The ensemble norm is defined as $||x|| = \sum_{i = 1}^{N} x_i/N$.
As shown in Fig.{\ref{Ksgs_budget_ensemble_sum}}, all unresolved source terms are dominant terms within the $0 \le t/\tau_{to} \le 3.0$.
Obviously, the SGS production term $- \tau_{ij} \widetilde{S_{ij}}$ is the most important term, considering the largest positive magnitude among the four source terms.
The ensemble  $P$ is positive, which represents the ensemble SGS kinetic energy forward scatter.
The ensemble SGS dilational dissipation rate $D_2$ is more than half of the ensemble SGS solenoidal dissipation rate $D_1$.
Compared with the incompressible turbulence system, the dilational dissipation rate cannot be neglected in supersonic turbulence.
The coarse-grained analysis on SGS dissipation rate for supersonic isotropic turbulence agrees with previous conclusion on compressible turbulence at a moderate turbulent Mach number ($Ma_t = 0.52$) \cite{martin2000subgrid}.
In addition, with the coarser grids, the ratio of the $D_2$ to $D_1$ becomes larger.
When modeling the SGS dissipation rate, the one-equation SGS model for compressible LES should consider the grids effect \cite{pomraning2002dynamic,park2007numerical,chai2012dynamic}.
The negative values of $\Pi$ represents the ensemble SGS pressure-dilation term  acts as the sink for SGS kinetic energy.
Different with the Fig.\ref{history_R1R2}, the SGS pressure-dilation term $\Pi$ doesn't change signs during the evolution and always preserves negative value on unresolved grids.
Especially, the magnitude of SGS pressure-dilation term $\Pi$ is in the order of unresolved SGS dissipation term within the acoustic time scale $\tau_a$, where acoustic time is defined as $\tau_a = Ma_t \tau_{to}$ \cite{hanifi2012transition}.
Thus, for decaying supersonic isotropic turbulence, it can be concluded that the SGS pressure-dilation term cannot be neglected as previous comments \cite{pomraning2002dynamic,wang2018kinetic}.
The literature for modeling SGS pressure-dilation term in subsonic regime can be found in Refs \cite{zeman1991decay,sarkar1992pressure,ristorcelli1997pseudo}, while it is still required to be studied for supersonic isotropic turbulence.
When $t/\tau_{to} \ge 3.0$, from Fig.{\ref{history_R1R2}}, the turbulent Mach number $Ma_t \approx 0.7$ and Taylor microscale Reynolds number
$Re_{\lambda} \le 20$, the resolved ensemble dissipation rate and pressure-dilation rate decrease to a small magnitude.
At the same time, on unresolved grids as Fig.{\ref{Ksgs_budget_ensemble_sum}}, the source terms decay to a very mall magnitude, which indicate that even the coarsest grids $A_3$ and $B_3$ are fine enough to resolve the flowfields.
This behavior is reasonable since the current system experience a very small Taylor microscale Reynolds number $Re_{\lambda} \le 20$.

\begin{figure}[!htp]
	\centering
	\includegraphics[width=0.32\textwidth]{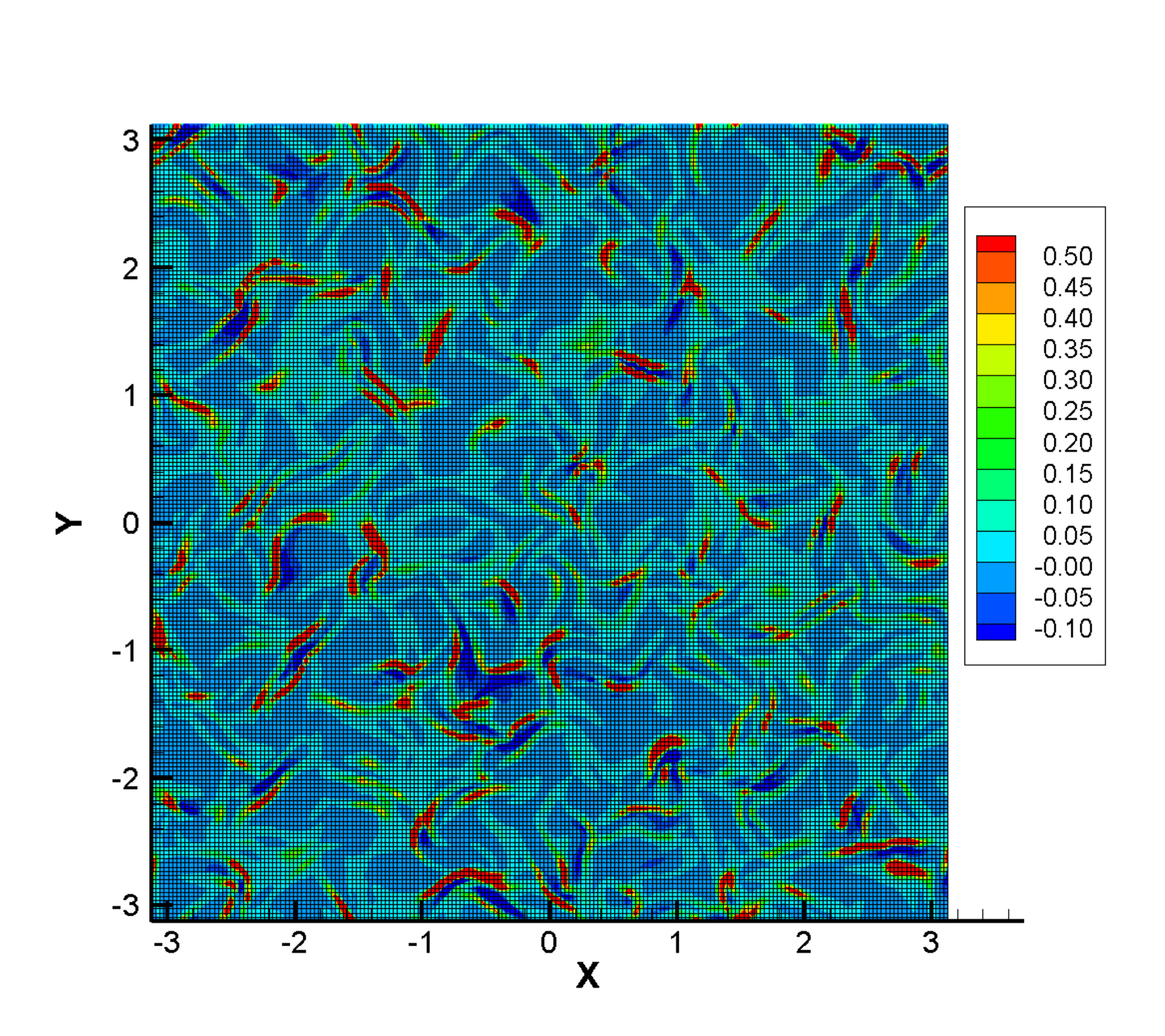}	
	\includegraphics[width=0.32\textwidth]{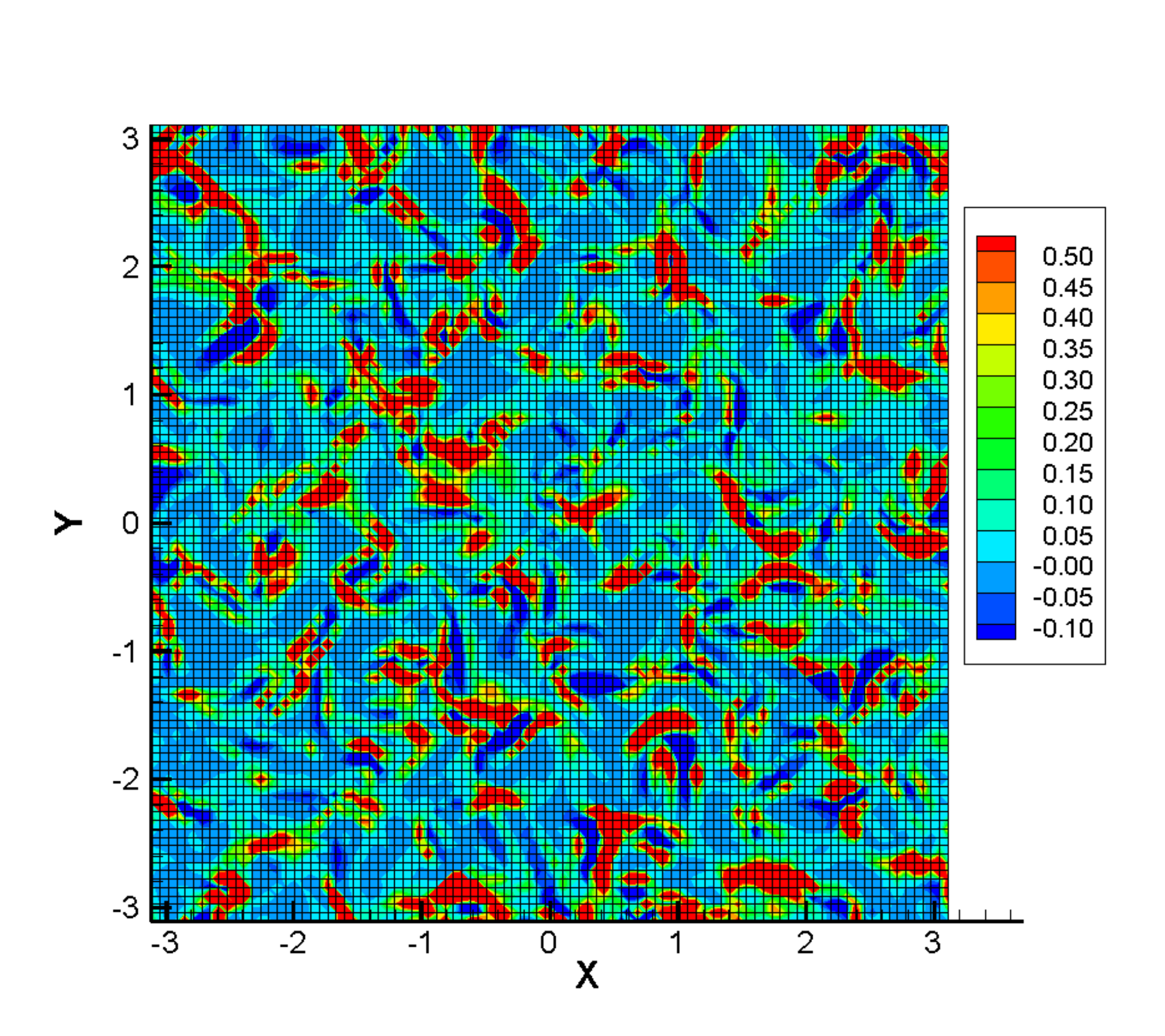}	
	\includegraphics[width=0.32\textwidth]{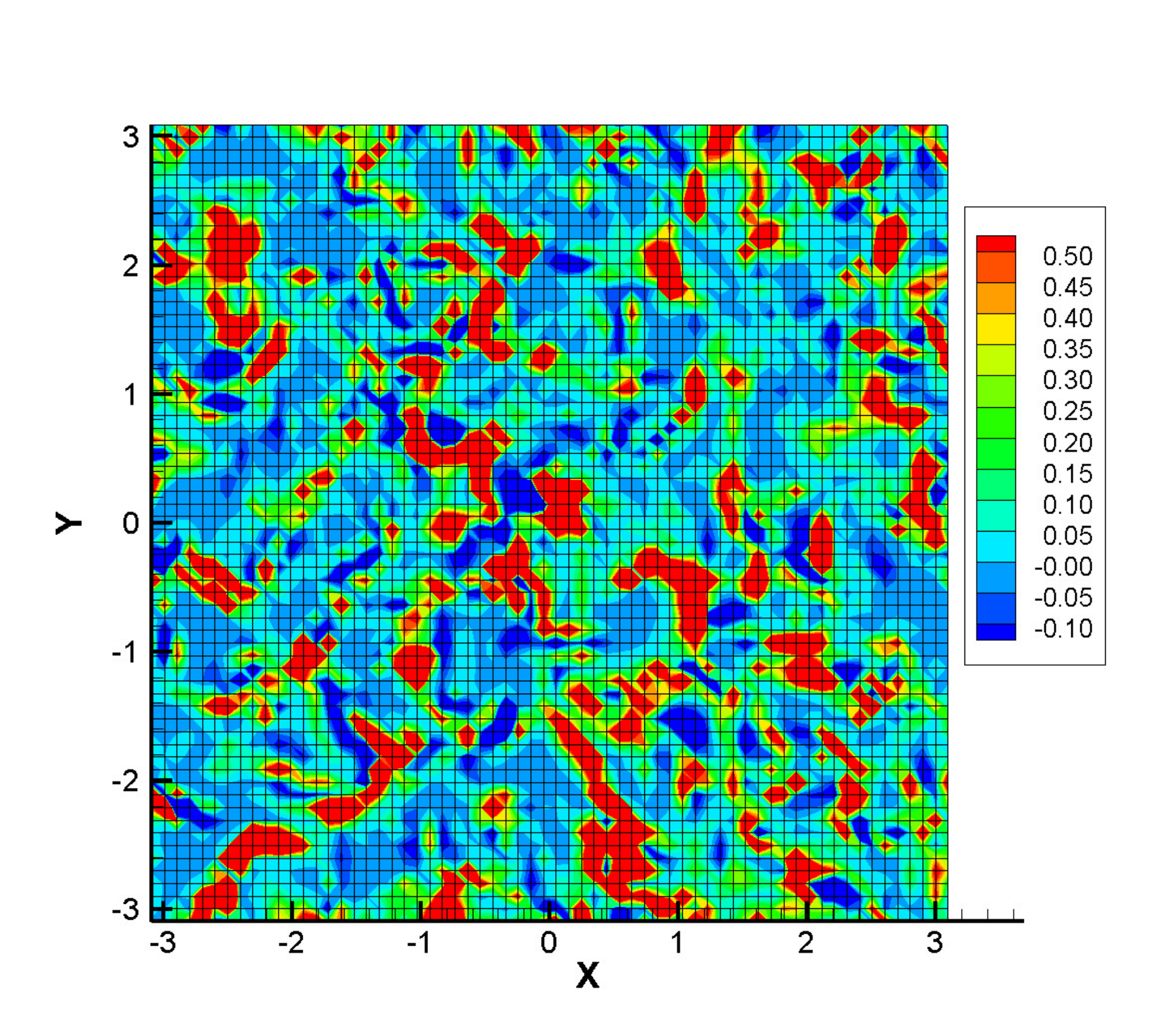}
	\includegraphics[width=0.32\textwidth]{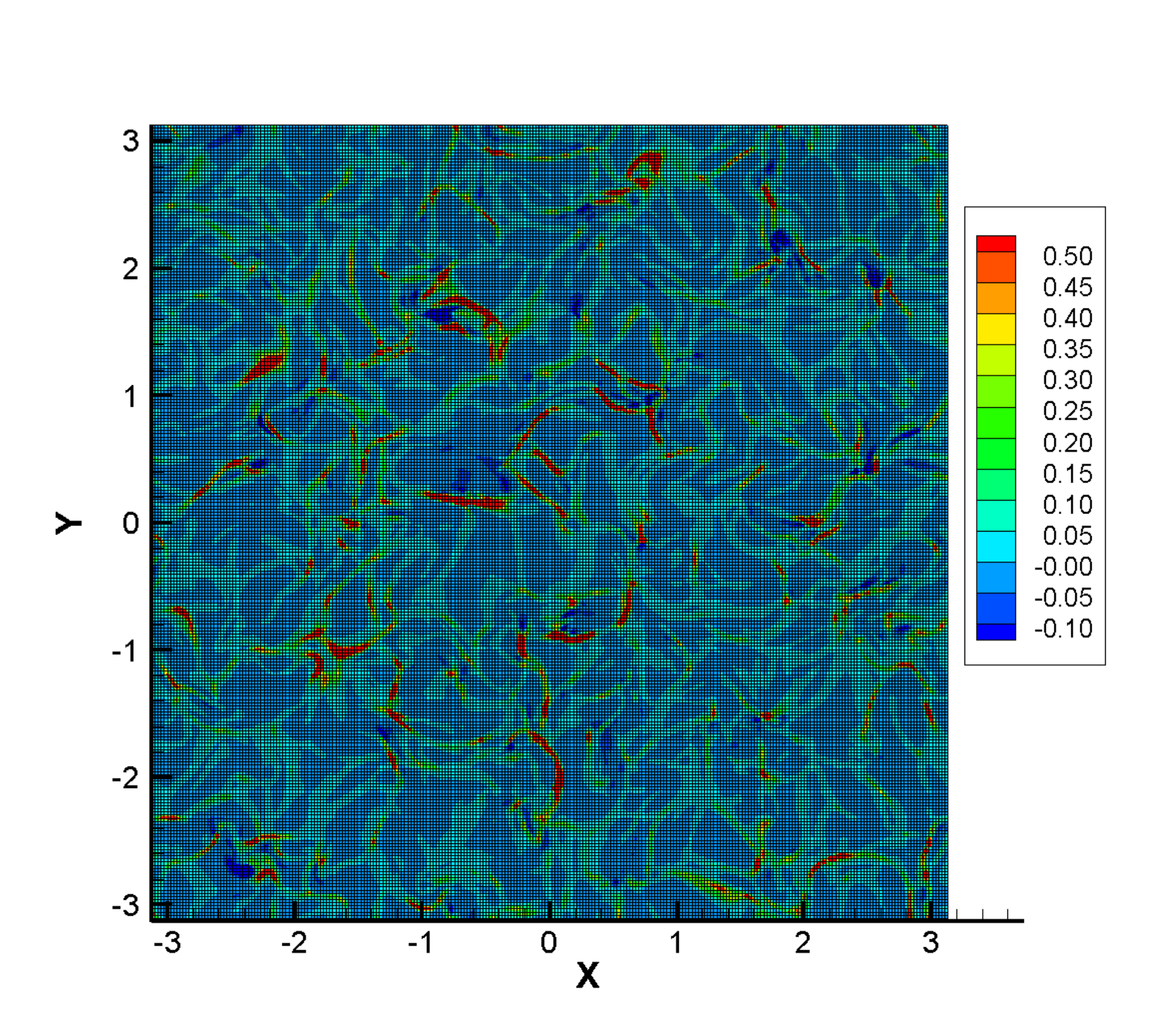}	
	\includegraphics[width=0.32\textwidth]{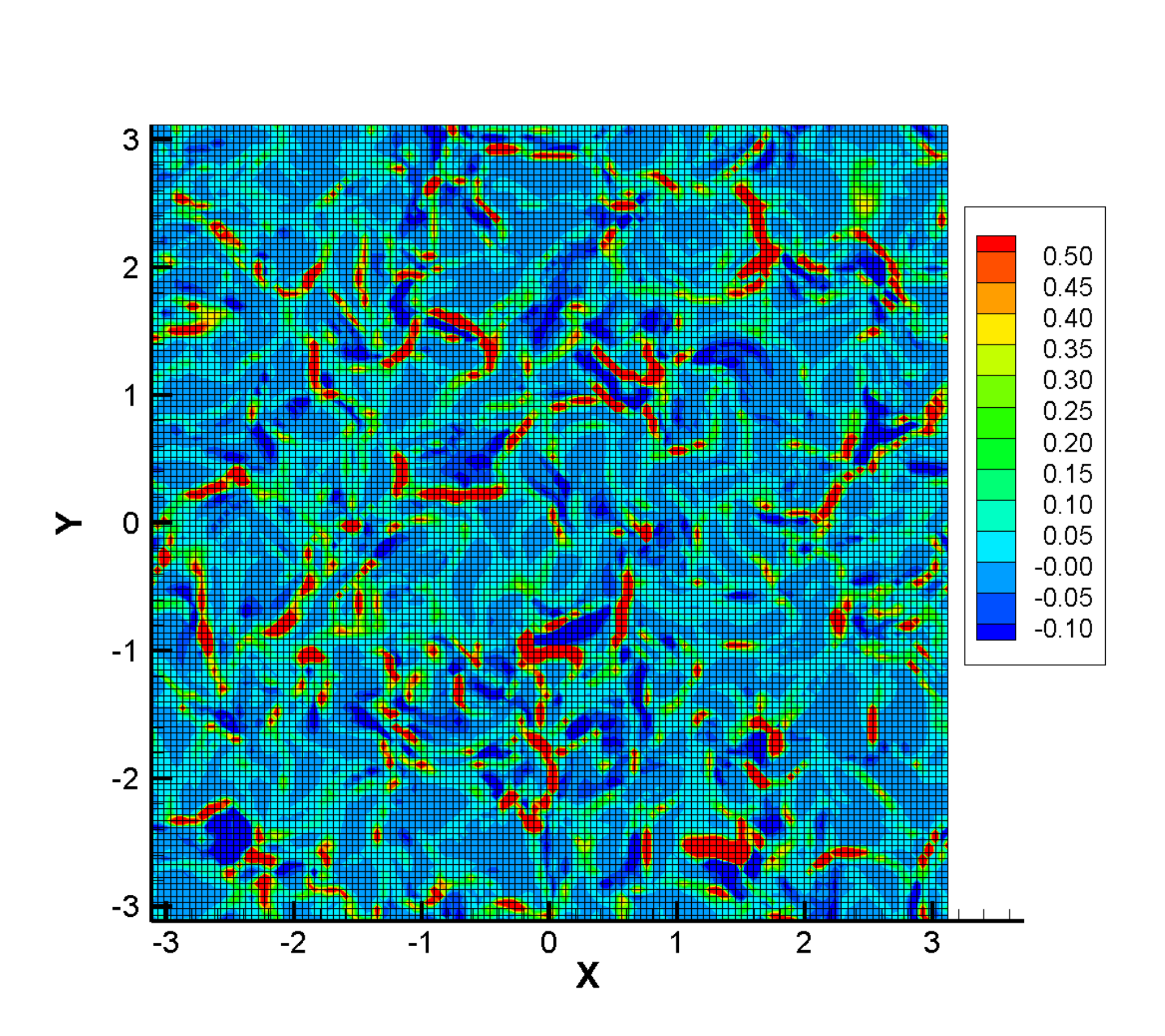}	
	\includegraphics[width=0.32\textwidth]{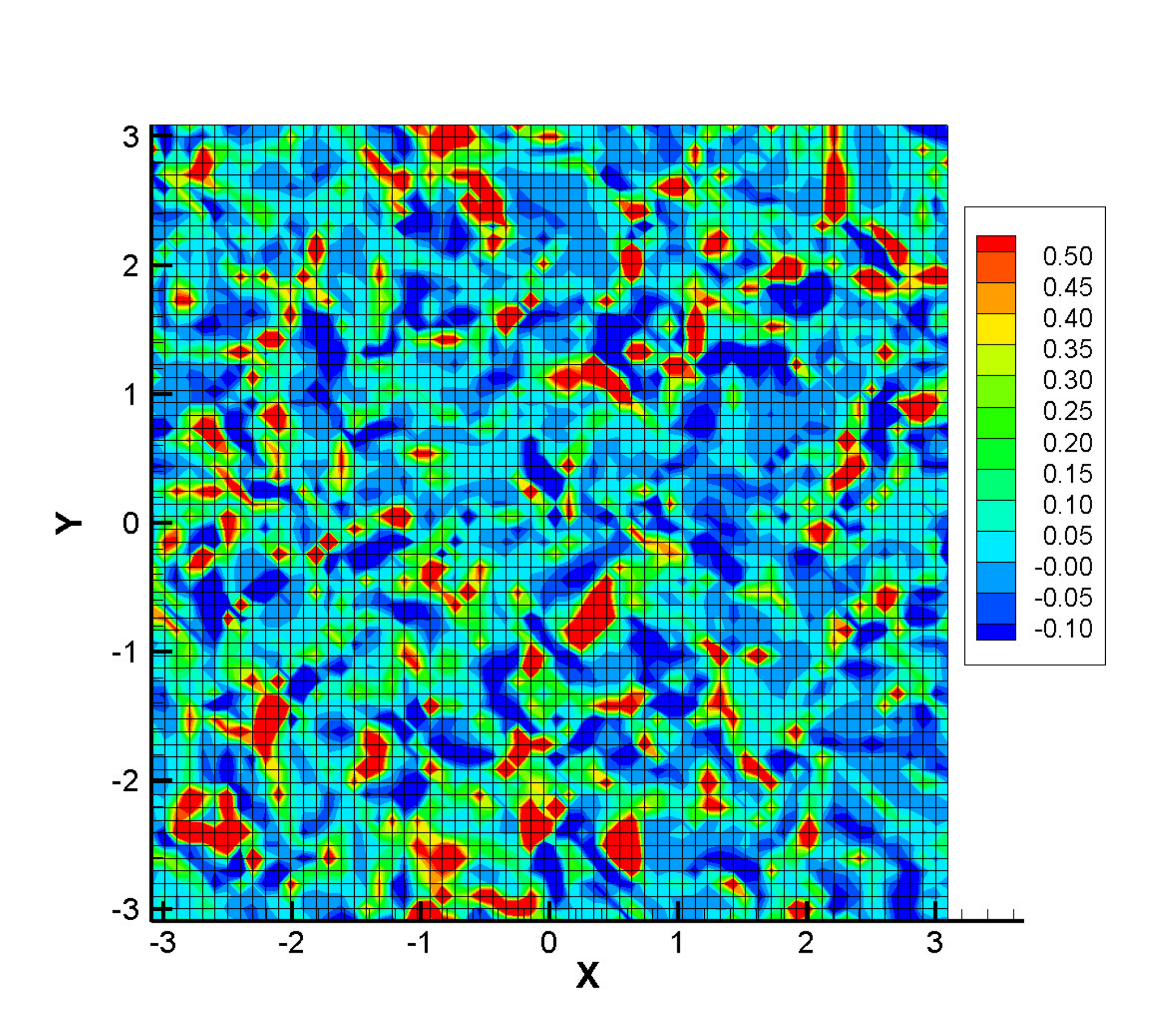}
	\vspace{-2mm}
	\caption{\label{t05_production}SGS production term $P$ for cases $A_1$, $A_2$ and $A_3$ at $t/\tau_{to} = 0.5$  at $z= 0$  (upper row), and cases $B_1$, $B_2$ and $B_3$ at $t/\tau_{to} = 1.0$ at $z= 0$  (lower row).}
\end{figure}
\begin{figure}[!h]
	\centering
	\includegraphics[width=0.45\textwidth]{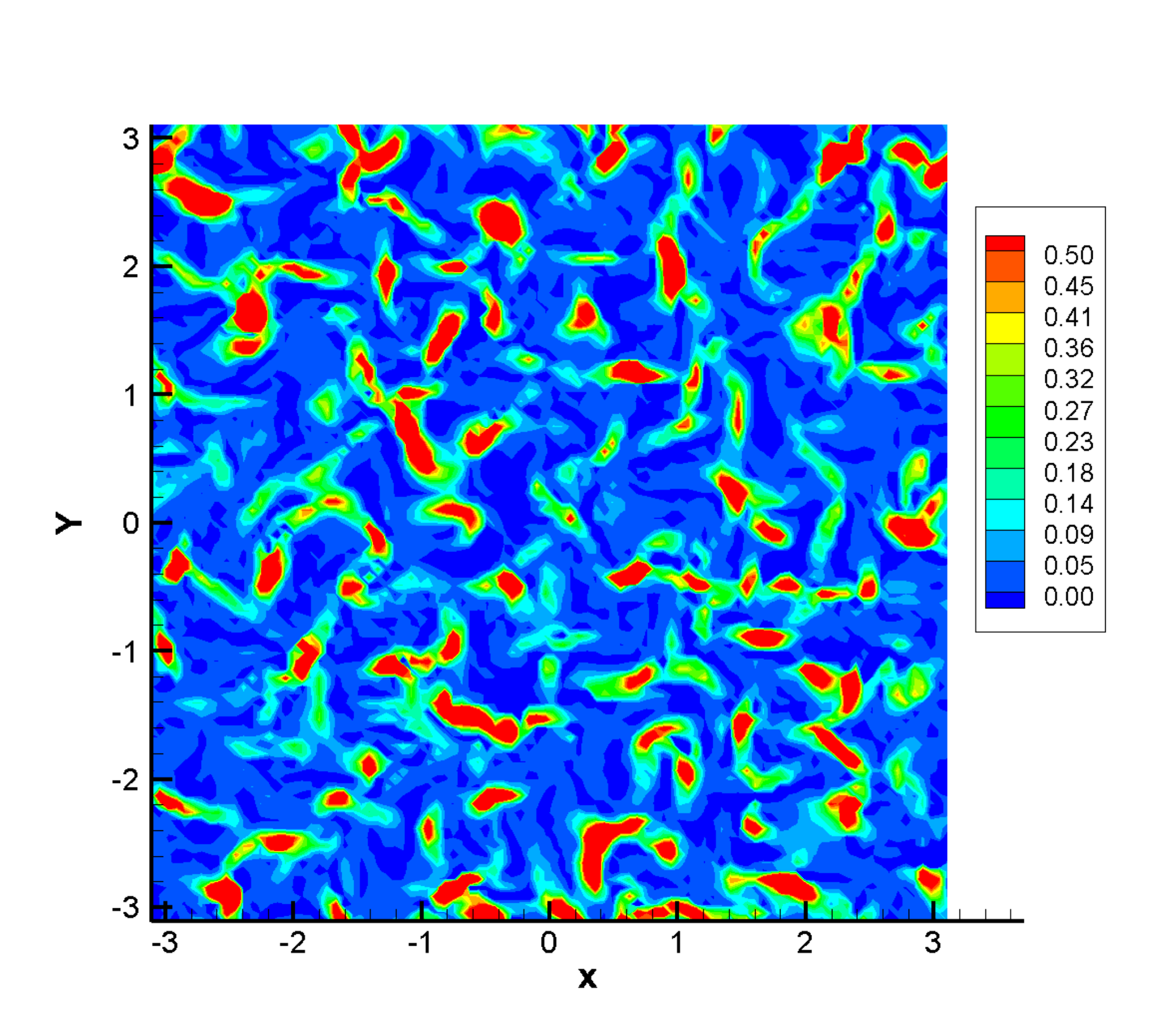}
	\includegraphics[width=0.45\textwidth]{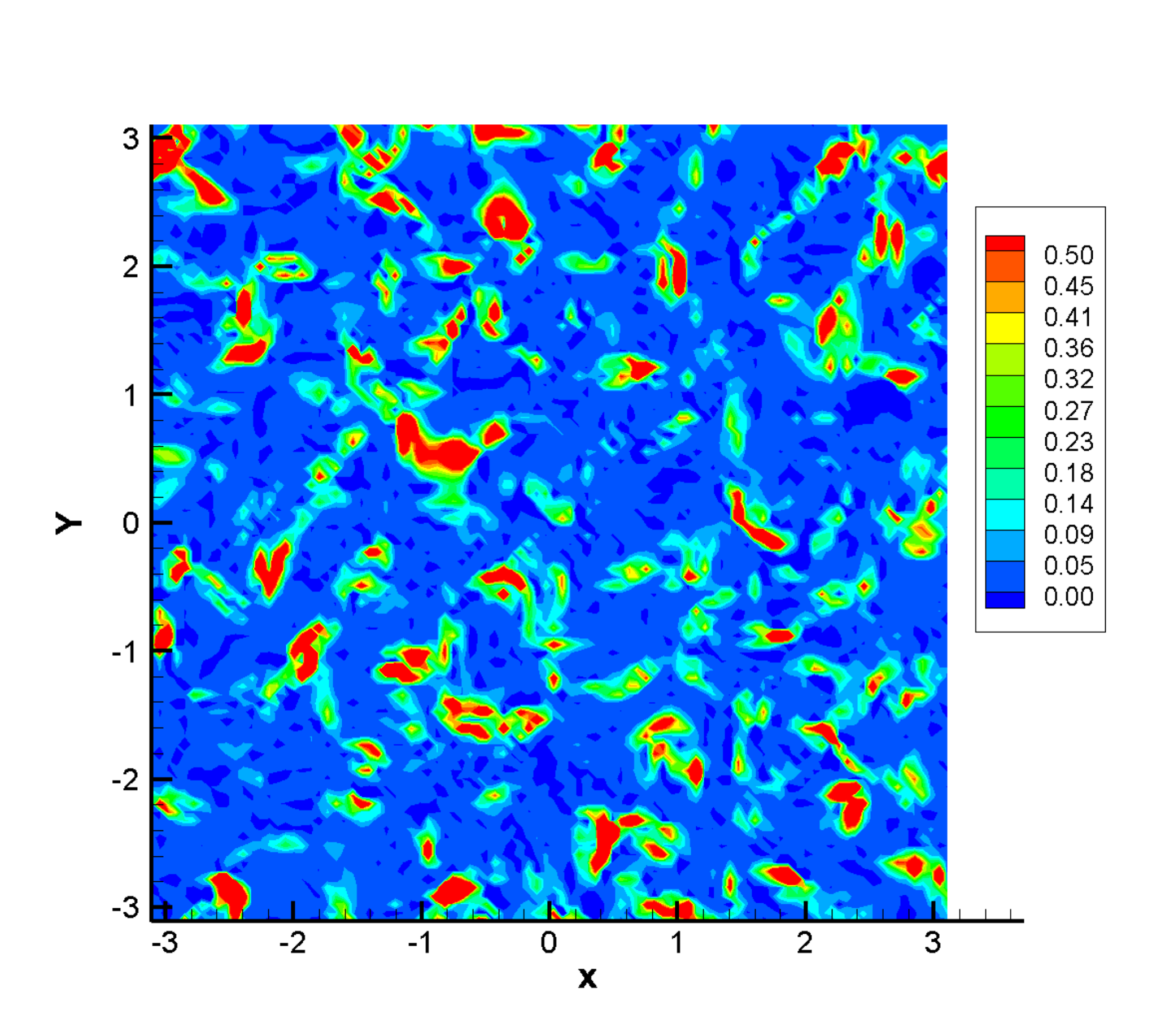}
	\vspace{-2mm}
	\caption{\label{t05_dissipation}SGS solenoidal dissipation term $D_1$ (left), SGS dilational dissipation term $D_2$ (right) for case $A_2$ at $t/\tau_{to} = 0.5$ at $z = 0$.}
	\centering
	\includegraphics[width=0.45\textwidth]{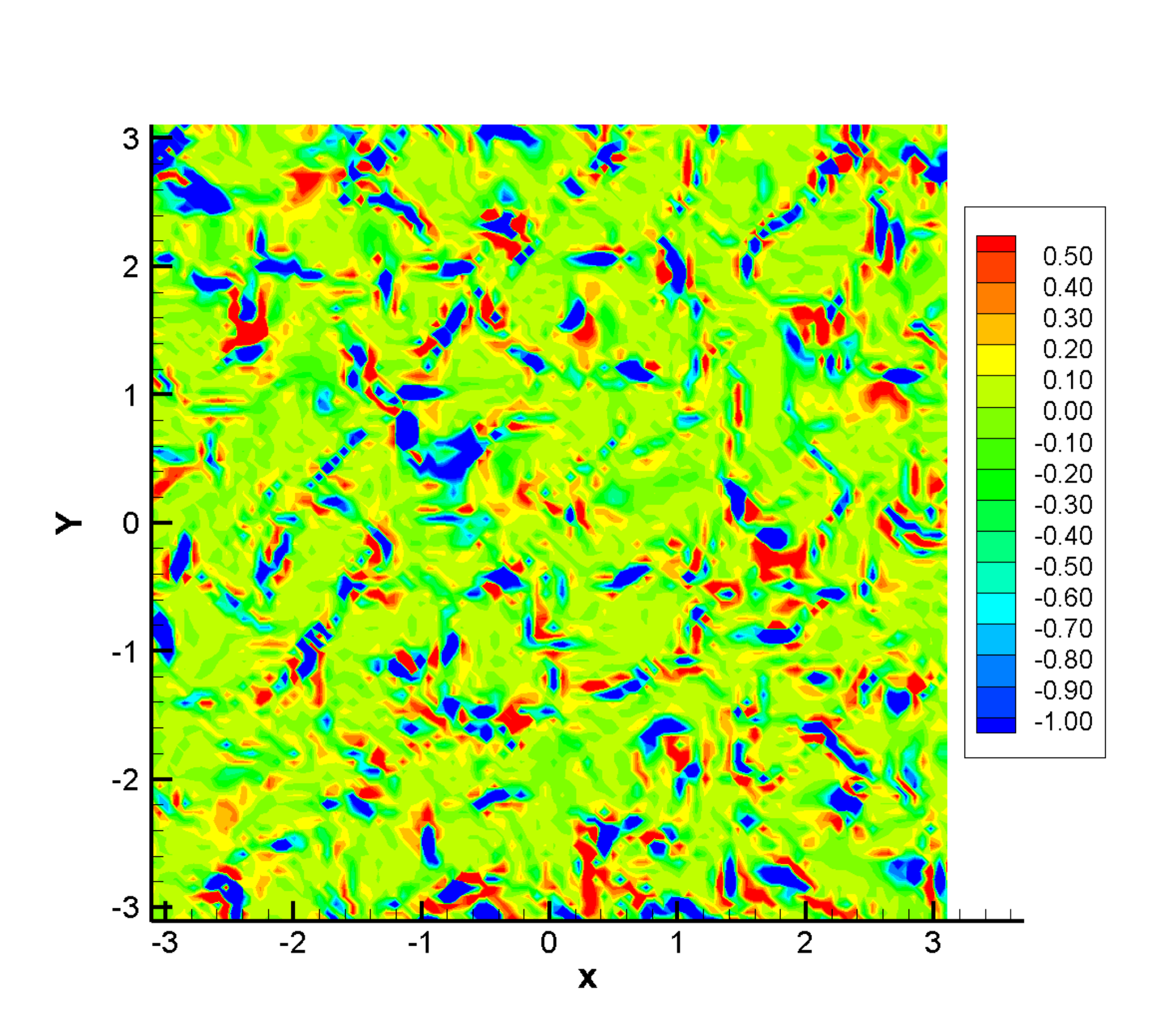}
	\includegraphics[width=0.45\textwidth]{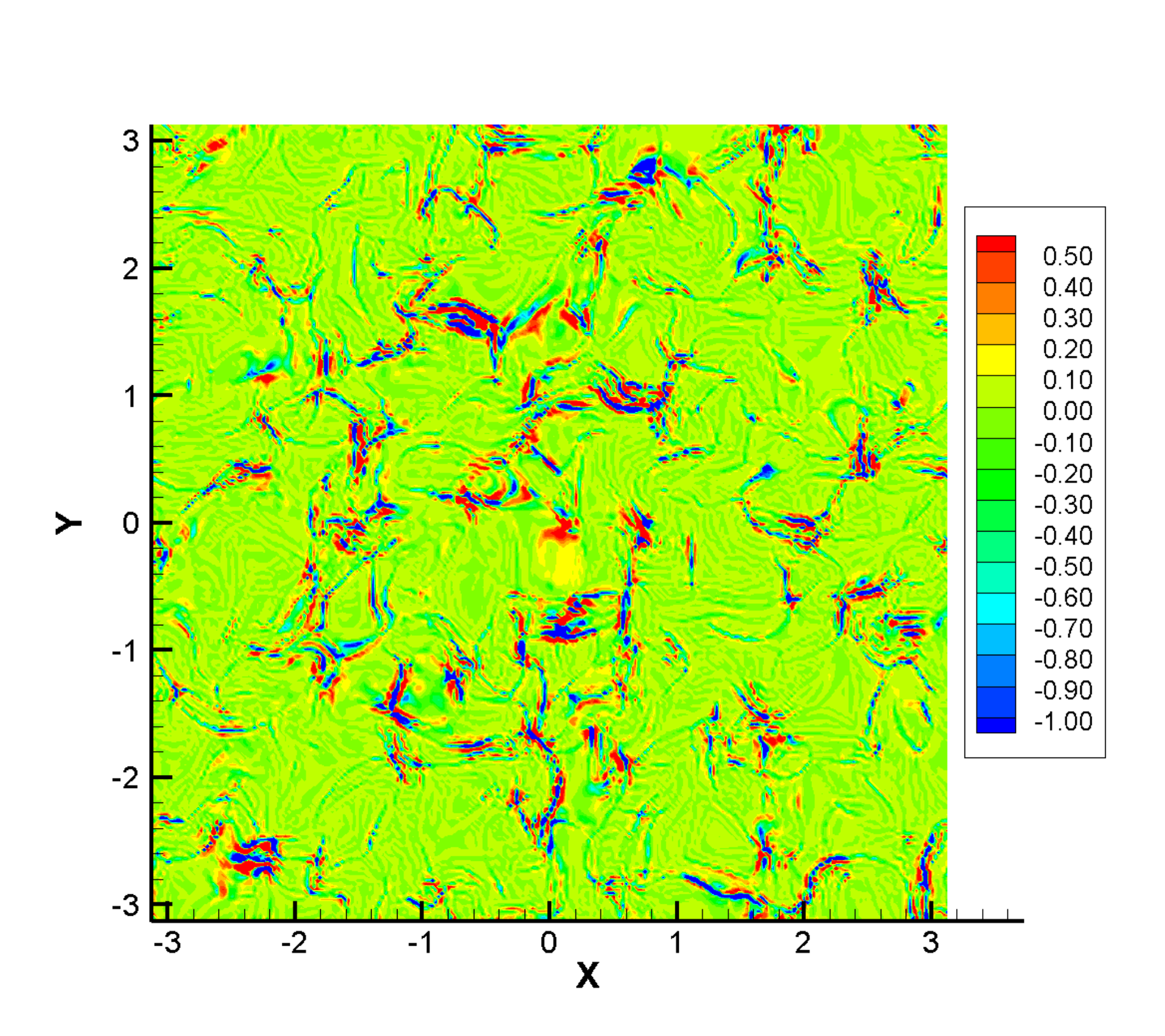}
	\vspace{-2mm}
	\caption{\label{t05_transfer} SGS pressure-dilation transfer term $\Pi$ for case $A_2$ at $t/\tau_{to} = 0.5$ (left) and $B_1$ at $t/\tau_{to} = 1.0$ (right) at $z = 0$.}
	\vspace{-2mm}
\end{figure}
The contours of SGS production term $P$ for cases $A_1$-$A_3$ at $t/\tau_{to} = 0.5$ and $B_1$-$B_3$ at $t/\tau_{to} = 1.0$ at $z= 0$ are presented in Fig.\ref{t05_production}.
The forward scatter and backscatter coexist \cite{piomelli1991subgrid,wang2018kinetic} and randomly distribute on the unresolved grids.
It can be seen that the magnitude and portion of positive $- \tau_{ij} \widetilde{S_{ij}}$ is larger than the negative ones, confirming that the ensemble forward scatter transfers the SGS turbulent kinetic energy from the resolved scales to the sub-grid scales.
To model the backscatter process in supersonic isotropic turbulence, the dynamic approach is recommended \cite{chai2012dynamic,moin1991dynamic}.
Contours of SGS solenoidal dissipation term $D_1$, dilational dissipation term $D_2$ for case $A_2$ at $t/\tau_{to} = 0.5$ and $B_1$ at $t/\tau_{to} = 1.0$ at $z= 0$ are presented as that in Fig.{\ref{t05_dissipation}}.
The dissipation rate is non-negative, and the high similarity between the $D_1$ and $D_2$ in spatial distribution are confirmed \cite{chai2012dynamic}.
Previous modeling \cite{sarkar1991analysis} on dilational dissipation rate $D_2 \propto Ma_t^2 D_1$ may still work in this supersonic isotropic turbulence, which will be studied in the following paper.
Figure {\ref{t05_transfer}} shows the SGS pressure-dilation transfer term $\Pi$ for case $A_2$ at $t/\tau_{to} = 0.5$ and $B_1$ at $t/\tau_{to} = 1.0$ at $z= 0$.
It can be seen that the magnitude and portion of negative $\Pi$ is larger than the positive ones, which confirms the ensemble SGS pressure-dilation term absorbing the $K_{sgs}$ as Fig.\ref{Ksgs_budget_ensemble_sum}.

\begin{figure}[!htp]
	\centering
	\includegraphics[width=0.32\textwidth]{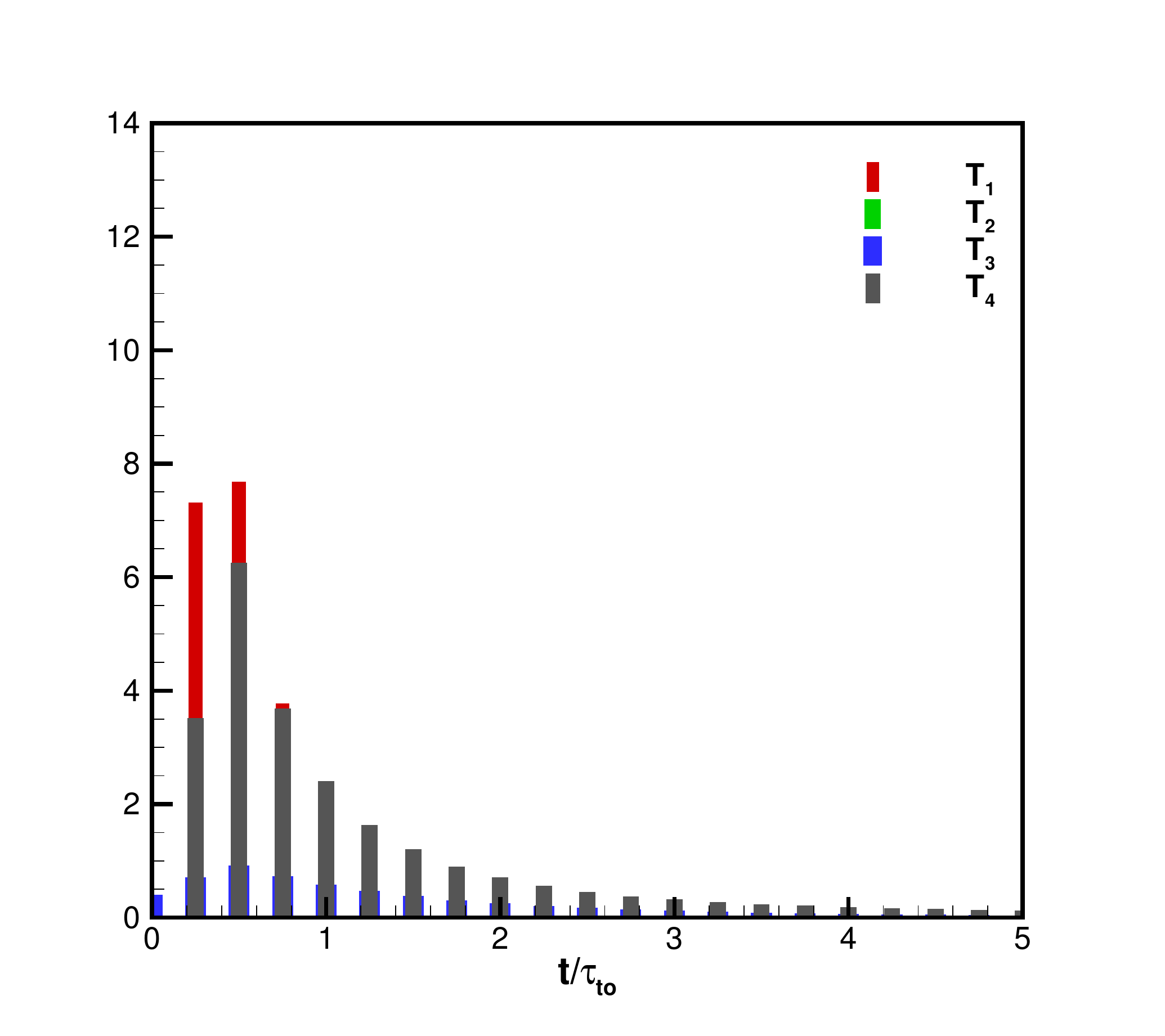}
	\includegraphics[width=0.32\textwidth]{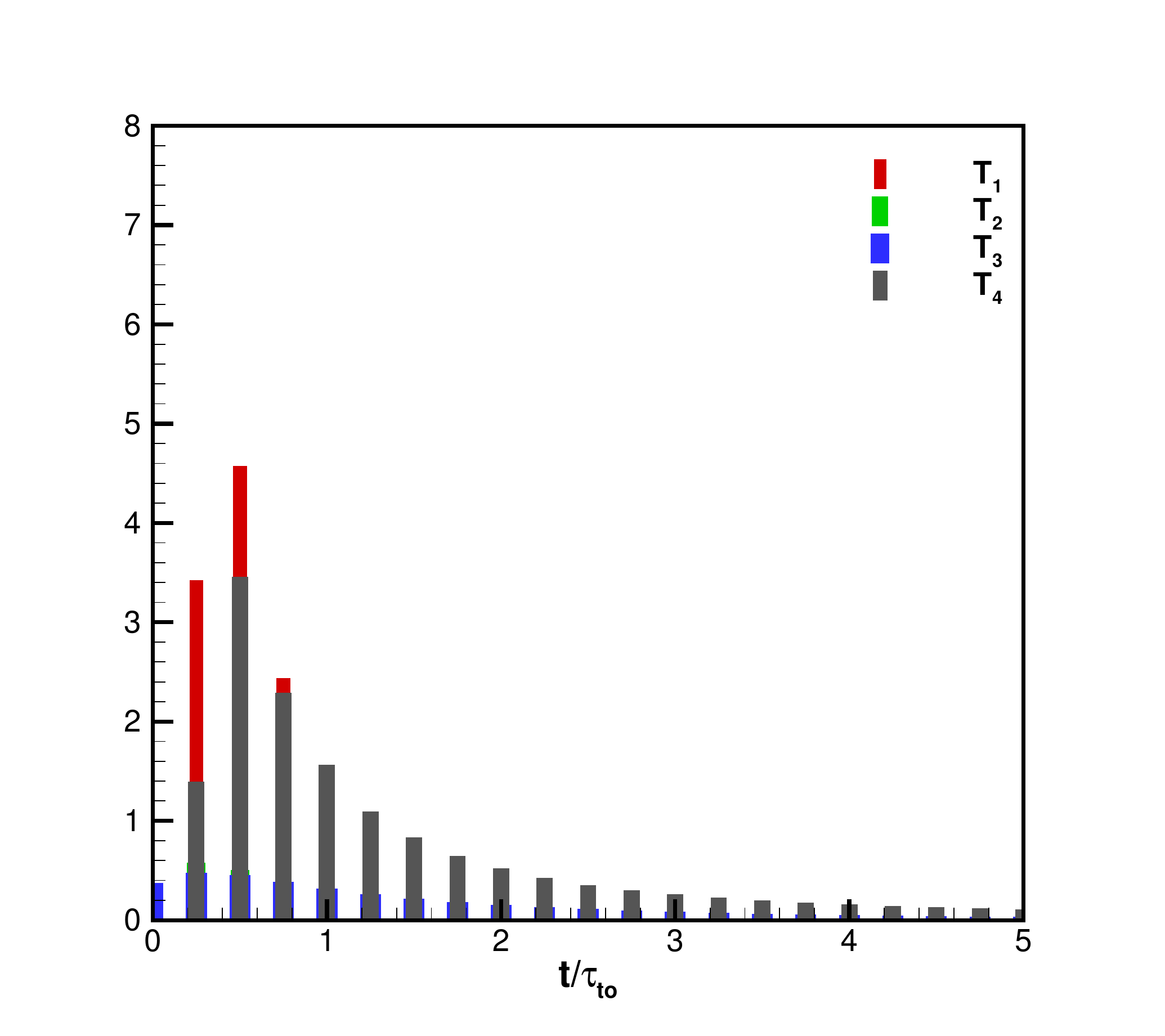}
	\includegraphics[width=0.32\textwidth]{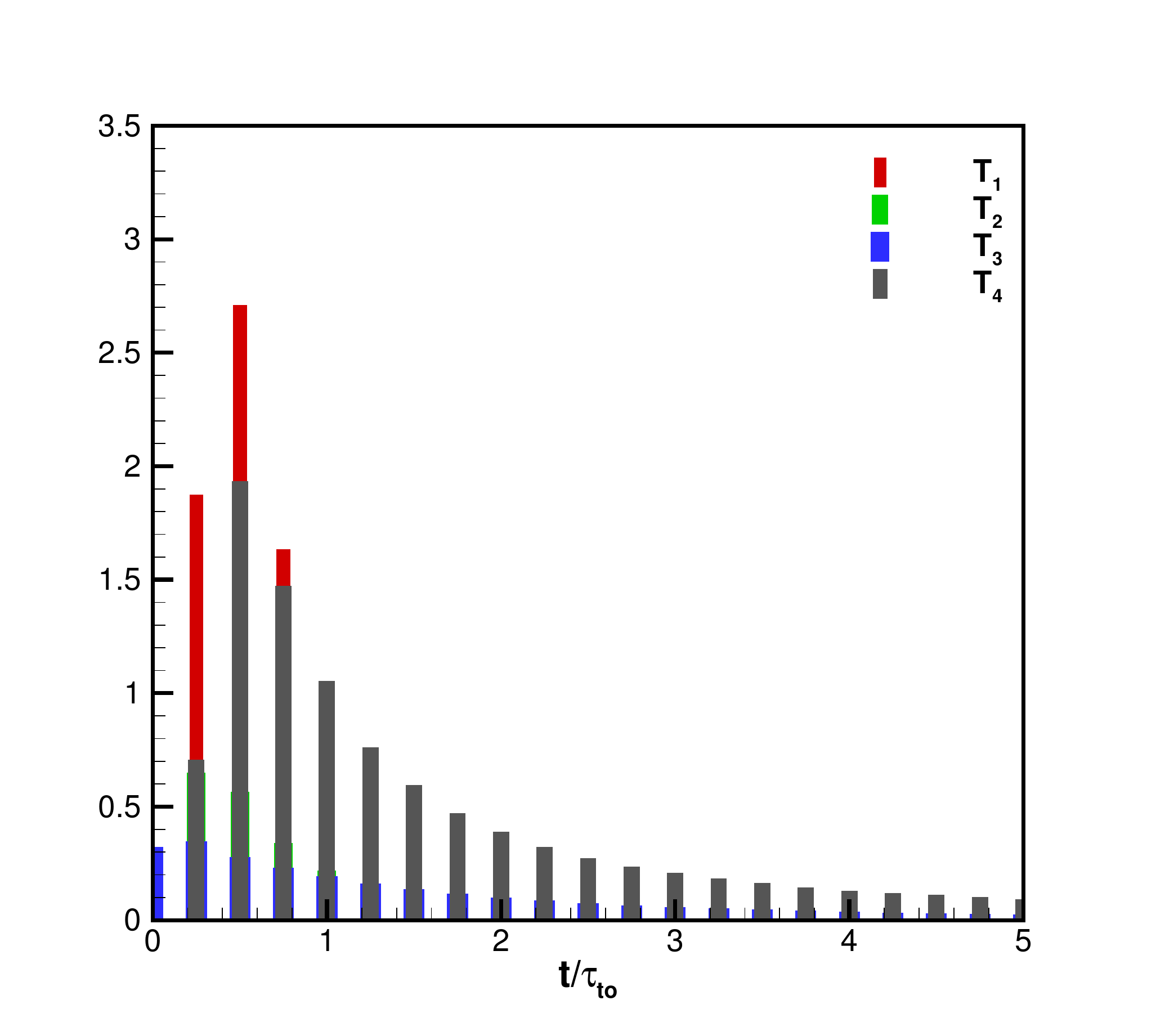}
	\includegraphics[width=0.32\textwidth]{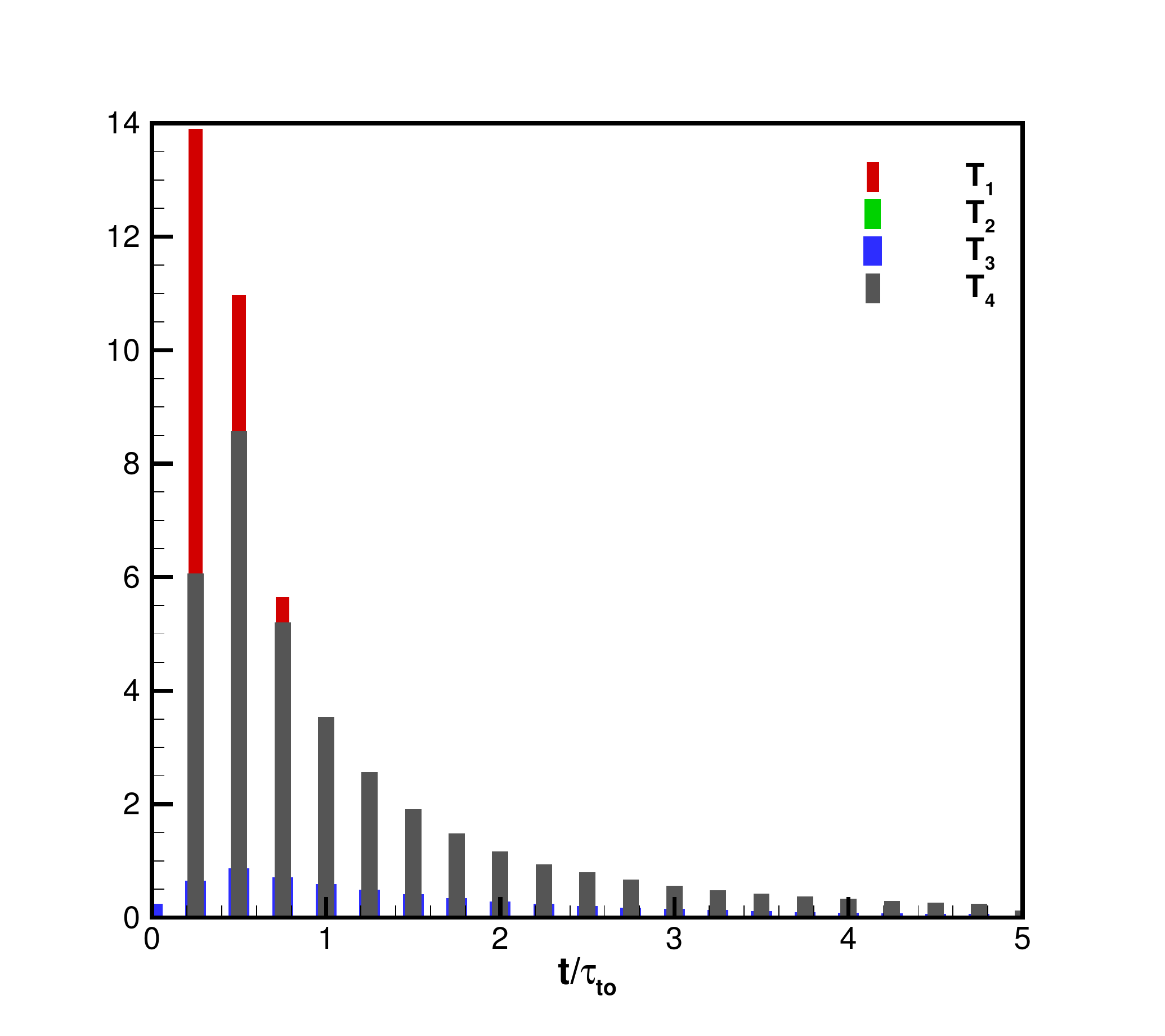}
	\includegraphics[width=0.32\textwidth]{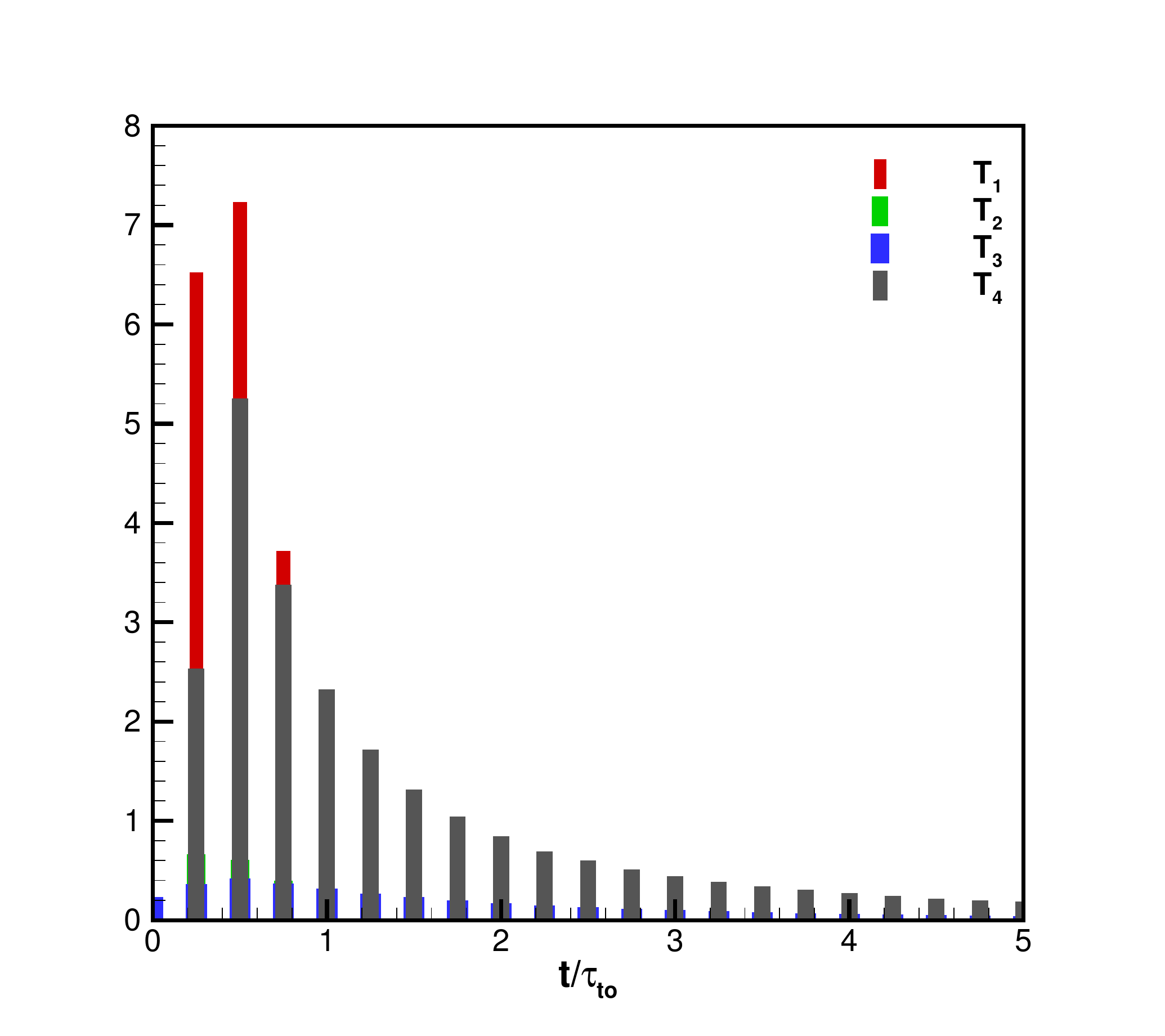}
	\includegraphics[width=0.32\textwidth]{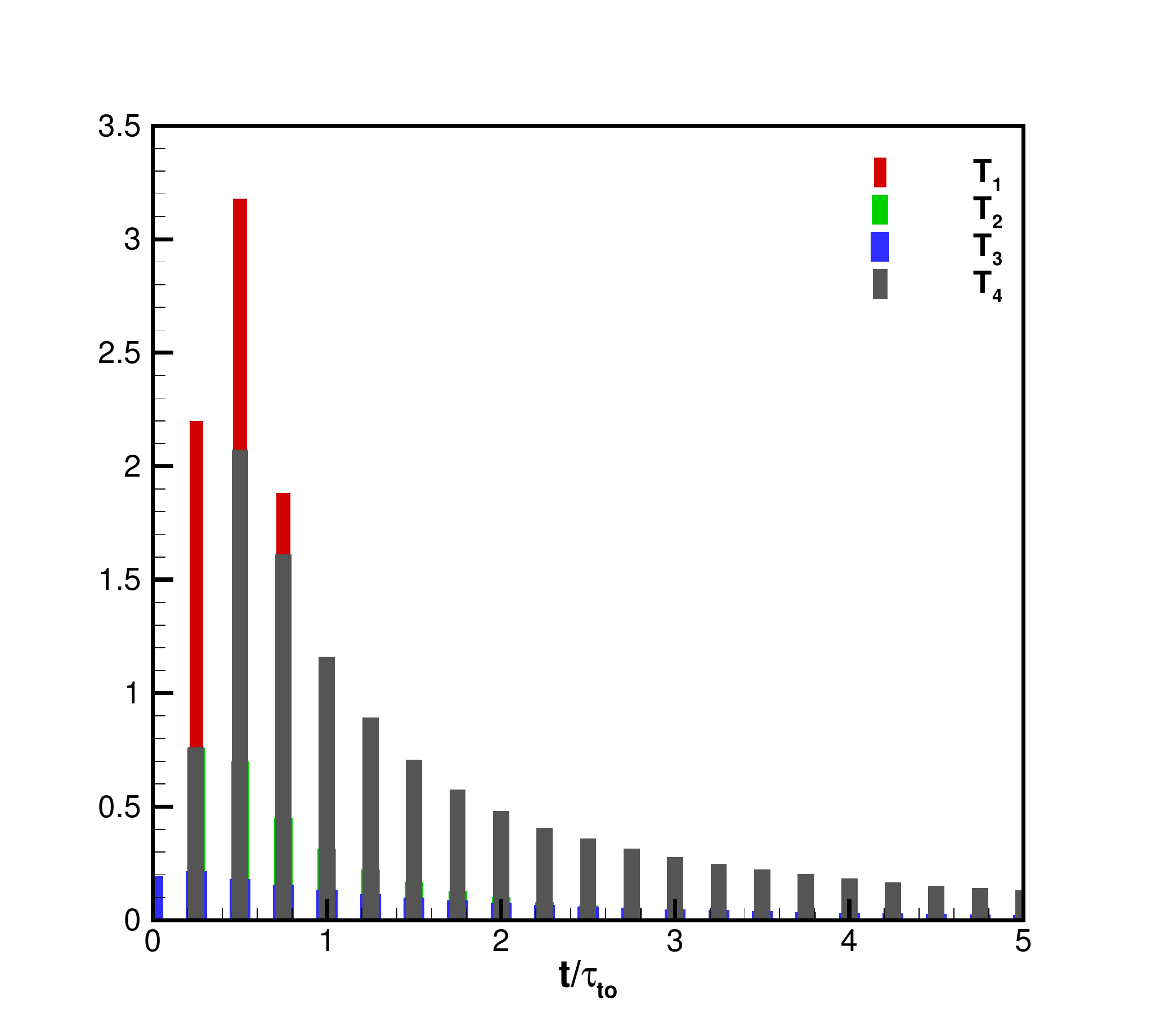}
	\vspace{-2mm}
	\caption{\label{Ksgs_budget_ensemble_transport}Coarse-grained budgets of SGS diffusion terms $T_1$, $T_2$, $T_3$ and $T_4$ for cases $A_1$, $A_2$, $A_3$ (upper row) and $B_1$, $B_2$, $B_3$ (lower row).}
	\vspace{-2mm}
\end{figure}
In previous study, the SGS diffusion terms are grouped and modeled together by the gradient-type models for both incompressible and compressible turbulent flows \cite{garnier2009large, wilcox1998turbulence}.
To study the delicate behavior of SGS diffusion terms in SGS kinetic energy equation, the coarse-grained analysis of dominant SGS diffusion terms is implemented.
Coarse-grained budget of SGS diffusion terms $T_1$, $T_2$, $T_3$ and $T_4$ for case $A_1$-$A_3$ and $B_1$-$B_3$ are presented as the Fig.{\ref{Ksgs_budget_ensemble_transport}}.
The budgets are computed in the $L_2$ norm, and the spatial derivatives are obtained by WENO-Z reconstruction as the Appendix B.
Because the ensemble of the sum of transport terms is equivalent to $0$, the $L_2$ norm is applied in analyzing the SGS diffusion terms.
The $L_2$ norm is defined as $||x||_{L_2} = (\sum_{i = 1}^{N} x_i^2)^{0.5}/N$.
As shown in Fig.{\ref{Ksgs_budget_ensemble_transport}}, within the $0 \le t/\tau_{to} \le 2.0$,  both the fluctuation velocity triple correlation term $T_1$ and the pressure-velocity correlation term $T_4$ are dominant terms.
$T_1$ and $T_4$ are about $10$ times larger than the negligible terms $T_2$ and $T_4$, i.e., $\|T_1\|_{L_2} \approx 10 \|T_4\|_{L_2}$.
The coarse-grained analysis on SGS diffusion terms for supersonic isotropic turbulence agrees with previous conclusion on subsonic isotropic turbulence \cite{vreman1995priori}, i.e., priori tests of a mixing layer up to Mach numbers $0.6$.
When $t/\tau_{to} \ge 3.0$, all SGS diffusion terms $T_1$-$T_4$ decay to a very mall magnitude similar as Fig.\ref{Ksgs_budget_ensemble_sum}.
This is because of the very small Taylor microscale Reynolds number $Re_{\lambda} \le 20$, even the coarsest grids $A_3$ and $B_3$ are fine enough to resolve the flowfields.

\begin{figure}[!h]
	\centering
	\includegraphics[width=0.45\textwidth]{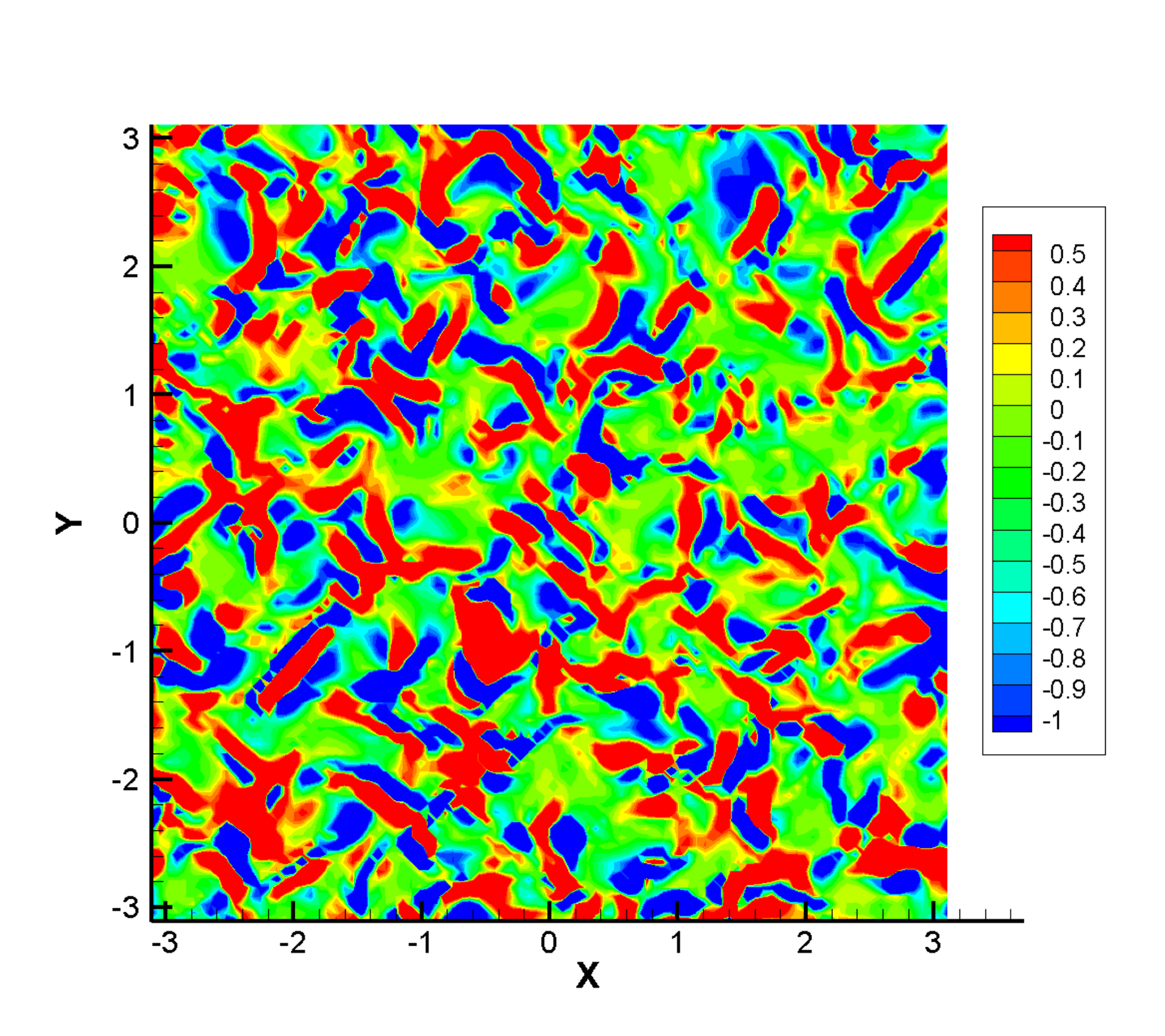}
	\includegraphics[width=0.45\textwidth]{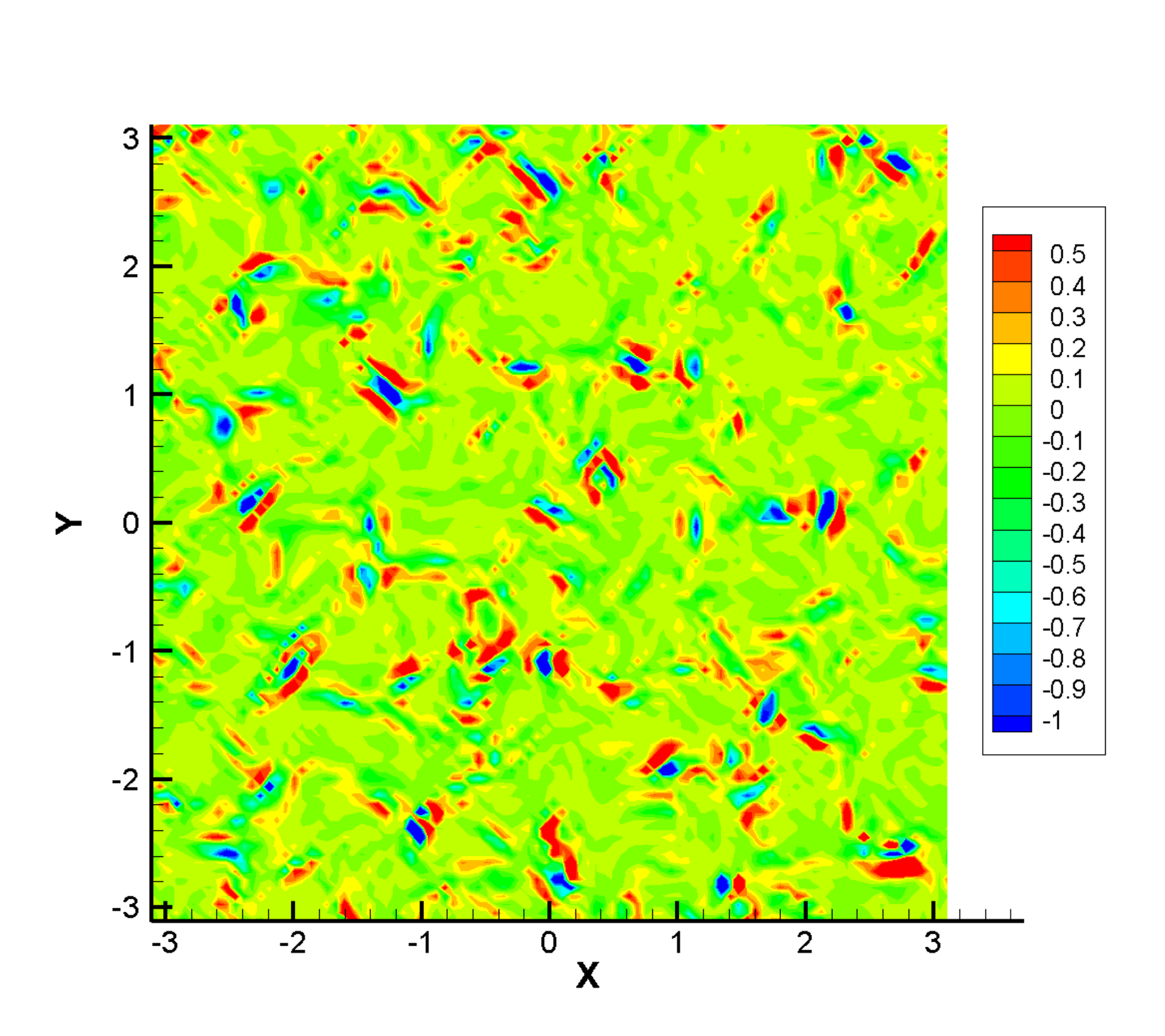} \\
	\includegraphics[width=0.45\textwidth]{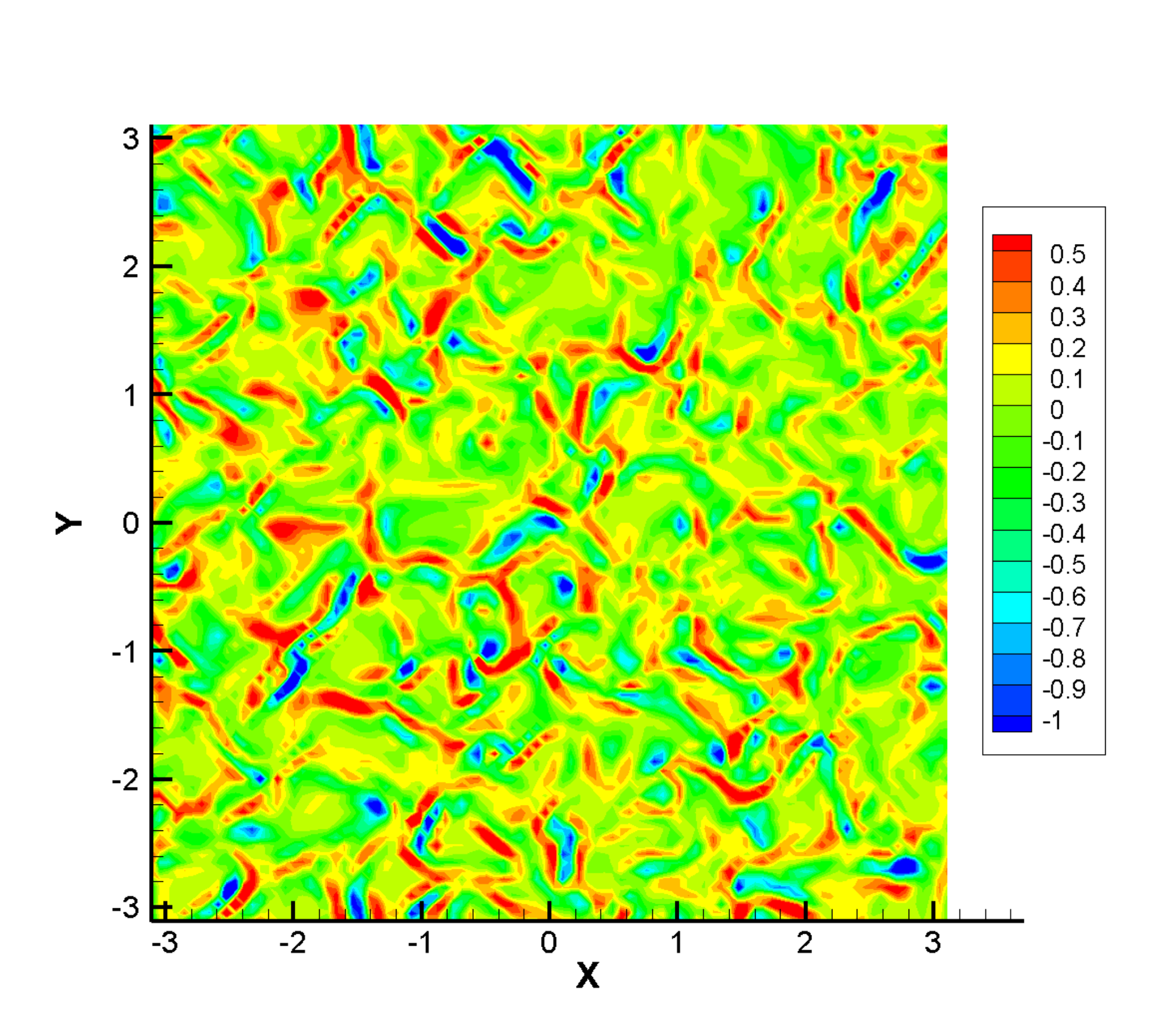}
	\includegraphics[width=0.45\textwidth]{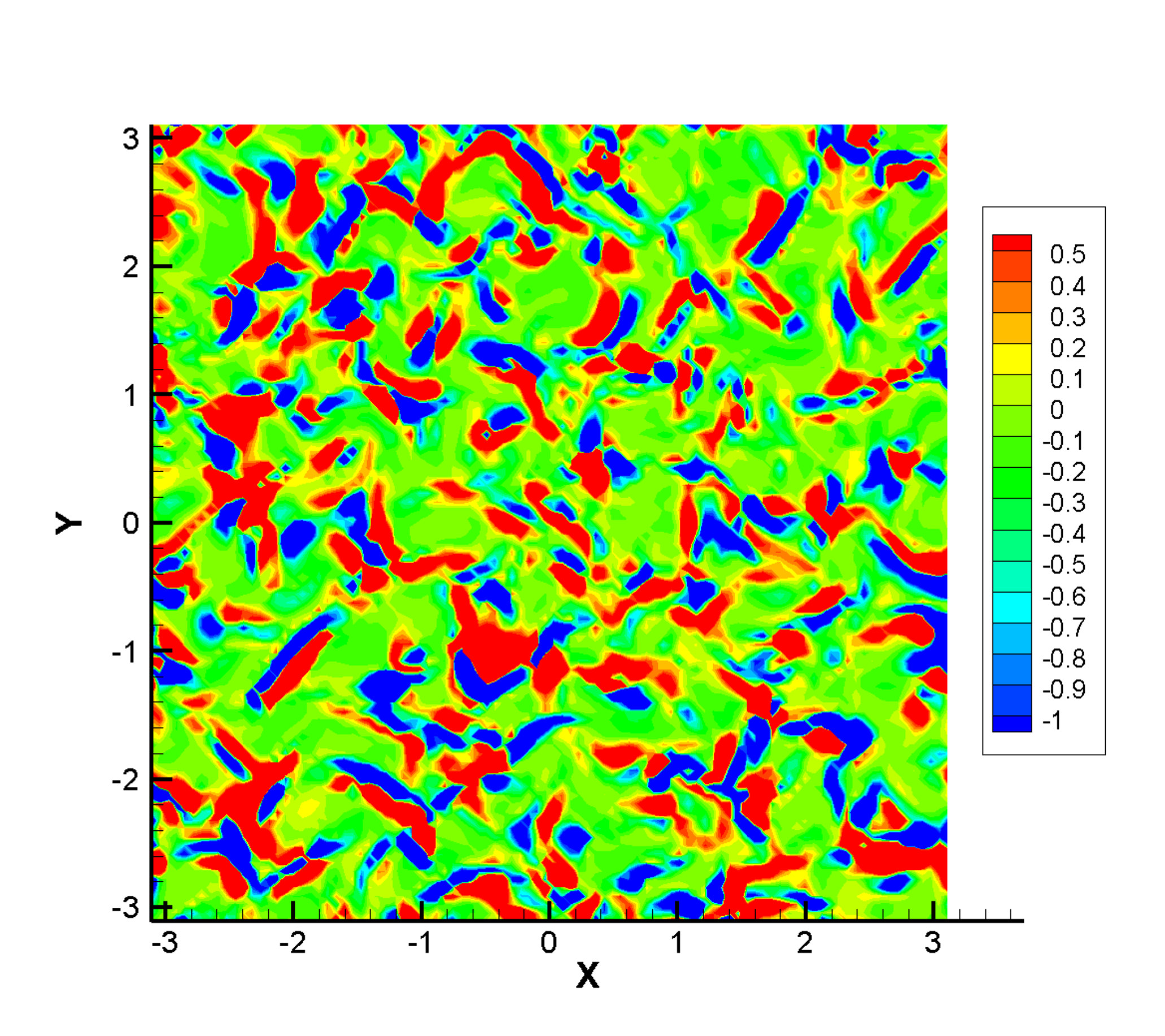}
	\vspace{-2mm}
	\caption{\label{t05_transport}SGS diffusion terms $T_1$, $T_2$ (upper row), $T_3$ and $T_4$ (lower row) diffusion terms for case $A_2$ at $t/\tau_{to} = 0.5$ at $z = 0$.}
	\vspace{-2mm}
\end{figure}
\begin{figure}[!h]
	\centering
	\includegraphics[width=0.45\textwidth]{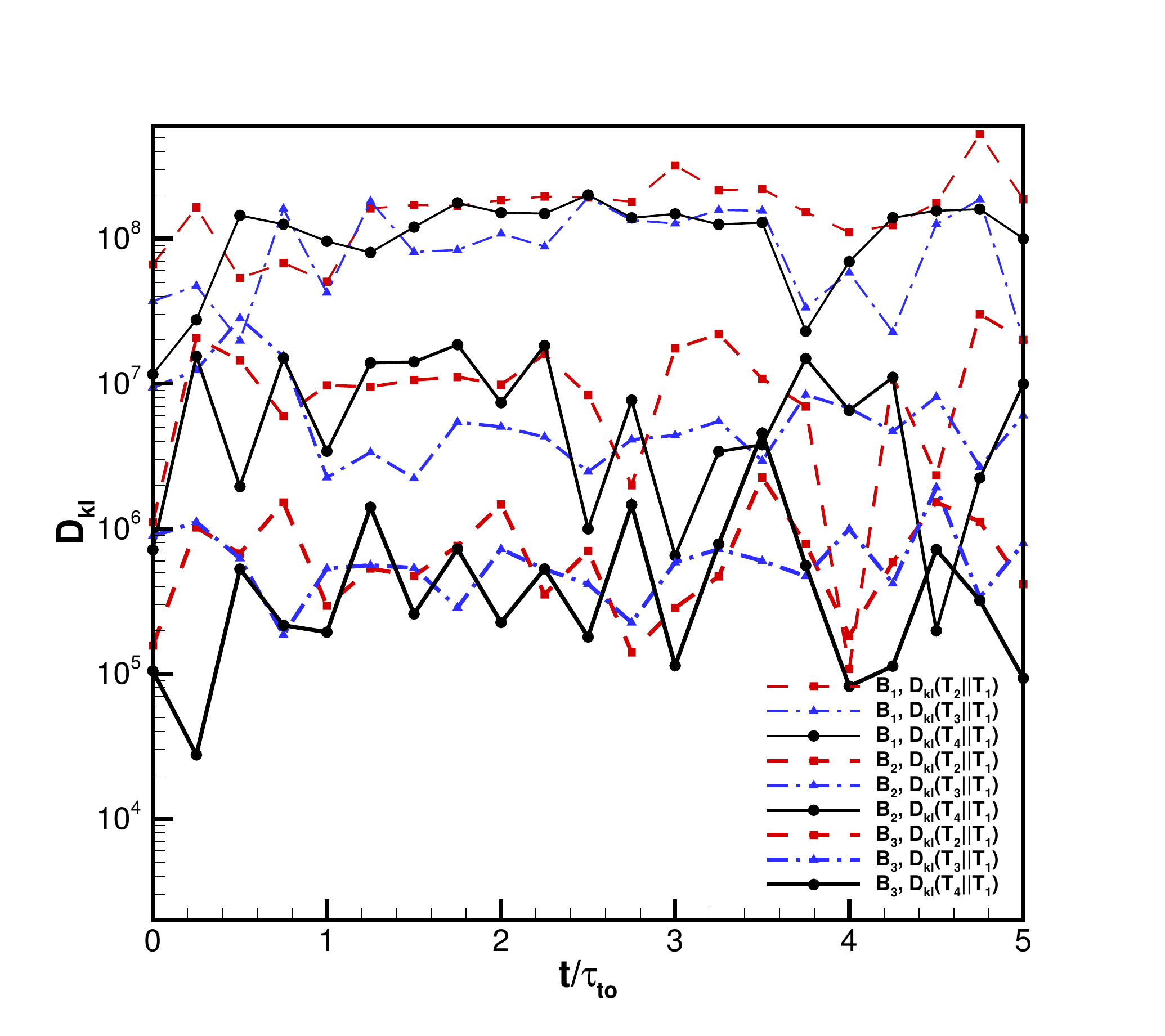}
	\includegraphics[width=0.45\textwidth]{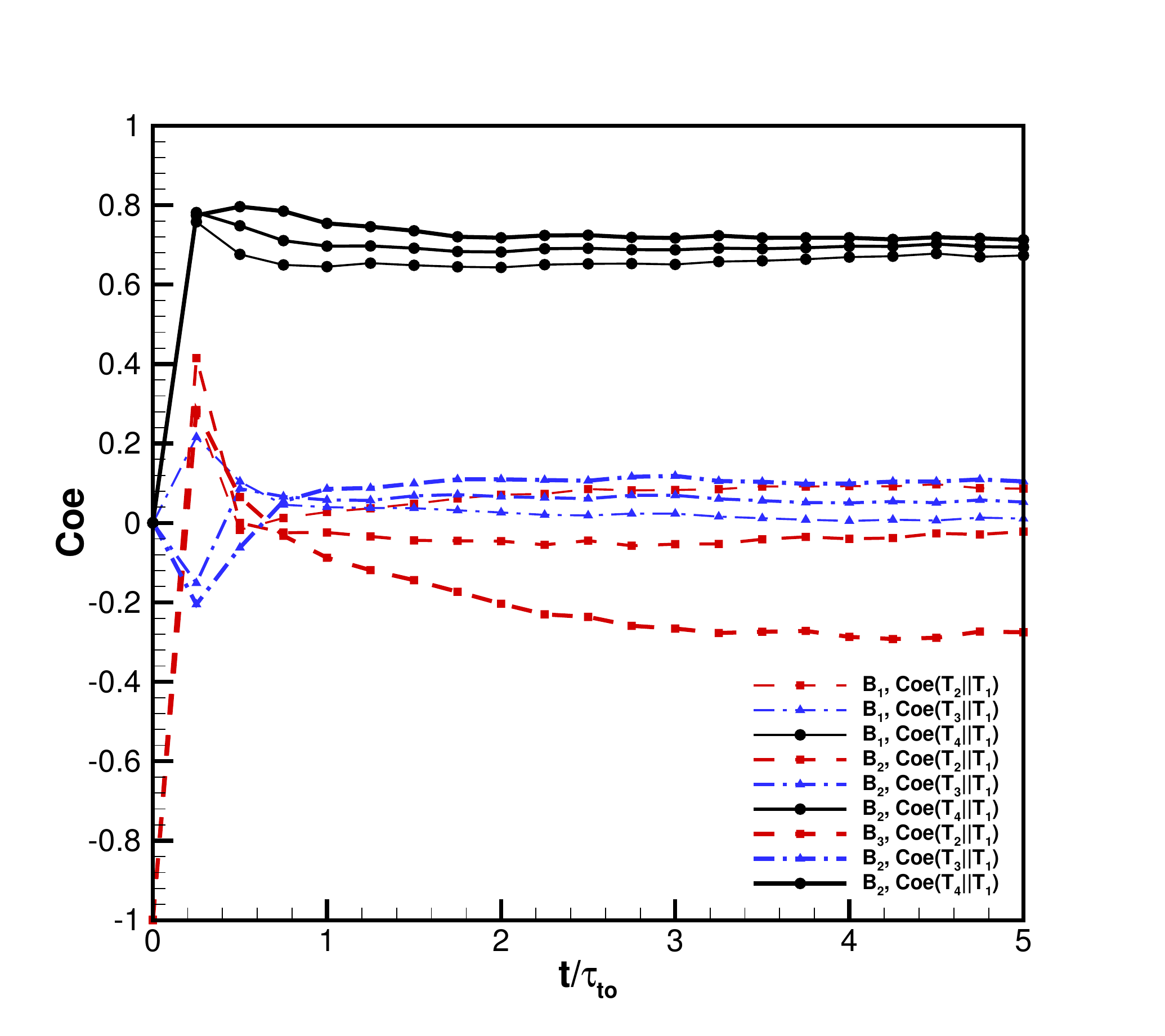}	
	\vspace{-2mm}
	\caption{\label{transport_kld_pdf}History of Kullback-Leibler divergence (left) and linear correlation coefficient (right) among the SGS diffusion terms $T_1$, $T_2$, $T_3$ and $T_4$ for cases $B_1$-$B_3$.}
	\vspace{-2mm}
\end{figure}

The contours of SGS diffusion terms $T_1$, $T_2$, $T_3$ and $T_4$ for case $A_2$ at $t/\tau_{to} = 0.5$  are presented in Fig.\ref{t05_transport},
in which the fluctuation velocity triple correlation term $T_1$ and the pressure-velocity correlation term $T_4$ behave more importantly than the SGS diffusion term $T_2$ and $T_3$.
To be of interest, the fluctuation velocity triple correlation term $T_1$ and the pressure-velocity correlation term $T_4$ are found to be highly correlated.
To further study the correlation, Kullback–Leibler divergence (KLD) \cite{kullback1951information} is introduced to measure the relationship of statistical behavior, namely, the correlation between two PDFs of SGS diffusion term.
In addition, the linear correlation coefficient is used to measure the spatial correlation of four SGS diffusion terms.
The KLD and linear correlation coefficient are defined as
\begin{equation}
\begin{aligned}\label{KL_divergence_correlation_coefficient}
  D_{kl}(\bm T_i||\bm T_1) &= \sum_{i} \bm T_i(i) \log {\frac{\bm T_i(i)}{\bm T_1(i)}}, \\
  Coe(\bm T_i||\bm T_1) &= \frac{\text{cov} (\bm T_i, \bm T_1)}{\sigma_{\bm T_i} {\sigma_{\bm T_1}}},
\end{aligned}
\end{equation}
where $\bm T_i$ is the PDF of SGS diffusion term $T_i$, and all PDFs of $T_i$  in this paper are obtained by dividing the SGS diffusion term range into $1000$ equivalent intervals.
$\text{cov}(\cdot, \cdot)$ represents the covariance of two random variables, and $\sigma_{\cdot}$ is  standard deviation of one random variables.
History of KLD and linear correlation coefficient among the SGS diffusion terms $T_1$, $T_2$, $T_3$ and $T_4$ for case $B_1$-$B_3$ are presented as Fig.{\ref{transport_kld_pdf}}.
The coarser grid is, the smaller magnitude of KLD is, indicating the closer relation between $T_i$ and $T_1$.
As different grids show different order of magnitude of KLD, it indicates that the grid effect should be considered for constructing the one-equation SGS model.
The linear correlation coefficient confirms the high correlation between $T_1$ and $T_4$, which indicates the strong coupling between the kinematics and thermodynamics in current supersonic isotropic turbulence.
When using the dynamic approach \cite{germano1991dynamic} to determine the dynamic coefficients for modeling SGS diffusion term \cite{chai2012dynamic}, both $T_1$ and $T_4$ should participate in the dynamic approach, instead of only considering $T_1$ as incompressible one-equation SGS model \cite{krajnovic2002mixed,de2008localized}.

\begin{table}[!h]
	\begin{center}
		\caption{\label{table_dominant_terms} Classification of terms in incompressible and compressible $K_{sgs}$ equation.}
		\vspace{3mm}
		\centering
		\begin{tabular}{c|c|c}
			\hline \hline
			Category              &Current compressible system              &Incompressible system    \\
			\hline
			Dominant terms        &$P$, $D_1$, $D_2$, $\Pi$, $T_1$, $T_4$   &$P$, $D_1$, $T_1$   \\
			\hline
			Negligible terms      &$T_2$, $T_3$                             &$D_2$, $\Pi$, $T_2$, $T_3$, $T_4$       \\
			\hline \hline
		\end{tabular}
	\end{center}
	\vspace{-5mm}
\end{table}
In summary, the classification of terms in the compressible $K_{sgs}$ equation are presented in Table.\ref{table_dominant_terms}.
Compared with incompressible turbulent system \cite{schumann1975subgrid}, current study points out the additional dominant terms $D_2$, $\Pi$ and $T_4$, which deserves further study for  high Mach number turbulence  modeling.
Compressible $K_{sgs}$ transport equation is analyzed, which paves the way for modeling the unknowns in compressible one-equation SGS model.
Subsequent paper will focus on the compressible one-equation SGS model for high turbulent Mach number turbulent flows.

\section{Conclusion}
In this paper, the coarse-grained analysis of compressible SGS turbulent kinetic energy budget $K_{sgs}$ is fully analyzed for constructing one-equation SGS model of compressible LES at high turbulent Mach number.
DNS on a much higher turbulent Mach number up to $Ma_{t} = 2.0$ has been obtained by HGKS, which provides the high-fidelity DNS data for coarse-grained analysis.
The exact compressible SGS turbulent kinetic energy $K_{sgs}$ transport equation is also derived with Favre filtering process.
Based on the compressible $K_{sgs}$ transport equation, the coarse-graining processes are implemented on unresolved grids.
The coarse-grained analysis of compressible $K_{sgs}$ budgets shows that all unresolved source terms are dominant terms, i.e., the SGS production term,
the SGS solenoidal dissipation term, the SGS dilational dissipation term, and the SGS pressure-dilation term. 
Especially, for the decaying supersonic isotropic turbulence, the SGS pressure-dilation term plays the significant role in SGS turbulent kinetic energy transfer, which cannot be neglected.
The coarse-grained analysis of SGS diffusion terms in compressible $K_{sgs}$ budgets shows both the fluctuation velocity correlation term and the pressure-velocity correlation term are dominant terms.
The pressure-velocity correlation term should participate in the dynamic approach when determining the dynamic coefficients for modeling SGS diffusion term.
The current coarse-grained analysis gives an indication of the order of magnitude of all unresolved terms in compressible $K_{sgs}$ budget, which provides a solid basis for compressible one-equation SGS model.
The compressible one-equation SGS model within the non-equilibrium time-relaxation kinetic framework for high turbulent Mach number turbulence will be presented in the subsequent paper.

\section*{Acknowledgement}
The current research is supported by National Science Foundation of
China (11701038, 11772281, 91852114), the Fundamental Research
Funds for the Central Universities (2018NTST19), and the National Numerical Windtunnel project. The authors would like to thank TianHe-II in Guangzhou for providing high performance computational
resources.

\section*{Appendix A: derivation of compressible $K_{sgs}$ transport equation}
For the filtering operator as Eq.(\ref{filtering}), the following two properties, namely linearity and commutation with differentiation \cite{pope2001turbulent} are required as
\begin{equation} \label{filtering_properties}
\begin{aligned}
\overline{\phi + \varphi} &= \overline{\phi} + \overline{\varphi},  \\
\overline{\frac{\partial \phi}{\partial s}} &= \frac{\partial \overline{\phi}}{\partial s},
\end{aligned}
\end{equation}
where $s = {\boldsymbol{x}}, t$. To avoid subgrid term appearing in the filtered continuity equation, the Favre filtering \cite{favre1965equations} as Eq.\eqref{favre_average} is considered.
For Favre filtering, only the linearity has been inherited as
\begin{equation} \label{filtering_Favre}
\begin{aligned}
    \widetilde{\phi + \varphi} &= \widetilde{\phi} + \widetilde{\varphi}.
\end{aligned}
\end{equation}
It should be noticed that the commutation with differentiation don't apply to the Favre filtering.
The SGS kinetic energy equation can be derived by subtracting the product of the Favre-filtered velocity and the filtered momentum equation from the filtered product of the velocity and momentum equation \cite{chai2012dynamic}
\begin{equation} \label{substraction_ns}
\begin{aligned}
\overline{U_i \times [(\rho U_i)_{,t} + (\rho U_i U_j)_{,j} + p_{,i} - (\sigma_{ij})_{,j}]} - \widetilde{U}_i \times \overline{(\rho U_i)_{,t} + (\rho U_i U_j)_{,j} + p_{,i} - (\sigma_{ij})_{,j}}  = 0,
\end{aligned}
\end{equation}
where $\rho$ is the density, $U_i$ is the velocity component, $p = \rho R T$ is the pressure, and $T$ is the temperature and $R$ is the gas constant.
Ignoring the bulk viscosity, the viscous stress $\sigma_{ij}$ is given by
\begin{equation*}
\begin{aligned}
\sigma_{ij} = \mu \big(U_{i,j} + U_{j,i} - \frac{2}{3}U_{k,k} \delta_{ij}\big),
\end{aligned}
\end{equation*}
where $\mu$ is the molecular viscosity, and $\delta_{ij}$ is the Kronecker symbol.
Based on properties of filtered process as Eq.(\ref{filtering_properties}) and Eq.(\ref{filtering_Favre}),
Eq.(\ref{substraction_ns})  can be rearranged term by term to derive SGS kinetic energy equation.

The first term $L_1$ is defined and grouped as
\begin{equation} \label{first_term}
\begin{aligned}
    L_1 = & \overline{U_i \times (\rho U_i)_{,t}} - \widetilde{U}_i \times \overline{(\rho U_i)_{,t}}
        = [\overline{\rho} (\widetilde{U_i U_i} - \widetilde{U}_i \widetilde{U}_i)]_{,t} - (\overline{\rho U_i U_{i,t}} - \overline{\rho} \widetilde{U}_i \widetilde{U}_{i,t}).
\end{aligned}
\end{equation}
The continuity and momentum equation can be used to replace $U_{i,t}$ as $\rho U_{i,t} = (\rho U_i)_{,t} - U_i \rho_{,t}$. Similarly, the filtered continuity equation and filtered momentum equation can be used to replace $\widetilde{U}_{i,t}$. Plugging above replacements into Eq.(\ref{first_term}), $L_1$ can be rewritten as
\begin{equation*} \label{first_term_2}
\begin{aligned}
    L_1 = & [\overline{\rho} (\widetilde{U_i U_i} - \widetilde{U}_i \widetilde{U}_i)]_{,t} + \overline{U_i \times [(\rho U_i     U_j)_{,j} + p_{,i} - (\sigma_{ij})_{,j}]} \\
        - & \widetilde{U}_i \times \overline{(\rho U_i U_j)_{,j} + p_{,i} - (\sigma_{ij})_{,j}} - [ \overline{U_i^2 (\rho U_j)_{,j}} - \widetilde{U}_i^2 (\overline{\rho} \widetilde{U}_j)_{,j}].
\end{aligned}
\end{equation*}
With the definition of SGS kinetic energy $\overline{\rho} (\widetilde{U_k U_k} - \widetilde{U}_k \widetilde{U}_k) = 2 \overline{\rho} K_{sgs}$ in Eq.(\ref{les_stress}), plugging $L_1$ into Eq.(\ref{substraction_ns}), leads to
\begin{equation} \label{substraction_ns_2}
\begin{aligned}
    2 (\overline{\rho} K_{sgs})_{,t} + 2 \left \{\overline{U_i \times [(\rho U_i     U_j)_{,j} + p_{,i} - (\sigma_{ij})_{,j}]} - \widetilde{U}_i \times \overline{(\rho U_i U_j)_{,j} + p_{,i} - (\sigma_{ij})_{,j}} \right \} \\
    = \overline{U_i^2 (\rho U_j)_{,j}} - \widetilde{U}_i^2 (\overline{\rho} \widetilde{U}_j)_{,j}.
\end{aligned}
\end{equation}
The second term $L_2$ can be defined and rewritten as
\begin{equation} \label{second_term}
\begin{aligned}
    L_2 = & \overline{U_i \times (\rho U_i U_j)_{,j}} - \widetilde{U}_i \times \overline{(\rho U_i U_j)_{,j}} \\
        = & \overline{(\rho U_i U_i U_j)_{,j} - \rho U_i U_j U_{i,j}} - \widetilde{U}_i \times [(\overline{\rho} \widetilde{U}_i \widetilde{U}_j)_{,j} + (\tau_{ij})_{,j}],\\
\end{aligned}
\end{equation}
where $\tau_{ij} = \overline{\rho} (\widetilde{U_i U_j} - \tilde{u}_i \tilde{u}_j)$ as defined in Eq.(\ref{les_stress}). Combining $L_2$ and the right-hand-side term in Eq.(\ref{substraction_ns_2}), we have
\begin{equation*} \label{second_term_2}
\begin{aligned}
    L_3 = & 2 \times L_2 - [\overline{U_i^2 (\rho U_j)_{,j}} - \widetilde{U}_i^2 (\overline{\rho} \widetilde{U}_j)_{,j}] \\
        = & 2 (\overline{\rho} K_{sgs} \widetilde{U}_j)_{, j} + [\overline{\rho}(\widetilde{U_i U_i U_j} - \widetilde{U_i U_i} \widetilde{U}_j)]_{,j} - 2 \widetilde{U}_i (\tau_{ij})_{,j}.
\end{aligned}
\end{equation*}
The last term in $L_3$ can be rewritten as
\begin{equation*} \label{second_term_3}
\begin{aligned}
    \widetilde{U}_i (\tau_{ij})_{,j} = (\tau_{ij} \widetilde{U}_i)_{,j} - \tau_{ij} \widetilde{U}_{i,j}
                                     = (\tau_{ij} \widetilde{U}_i)_{,j} - \tau_{ij} \widetilde{S}_{ij},
\end{aligned}
\end{equation*}
where the decomposition $\widetilde{U}_{i,j} = \widetilde{S}_{ij} + \widetilde{\Omega}_{ij}$ is involved, $\widetilde{S}_{ij} = (\widetilde{U}_{i,j} + \widetilde{U}_{j,i})/2$ and $\widetilde{\Omega}_{ij} = (\widetilde{U}_{i,j} - \widetilde{U}_{j,i})/2$. $\tau_{ij}  \widetilde{\Omega}_{ij} = 0$ because it involves multiplication of a symmetric tensor $\tau_{ij}$ by an anti-symmetric tensor $\widetilde{\Omega}_{ij}$. Plug $L_3$ into Eq.(\ref{substraction_ns_2}), which leads to
\begin{equation} \label{substraction_ns_3}
\begin{aligned}
    2 (\overline{\rho} K_{sgs})_{,t} + 2 (\overline{\rho} K_{sgs} \widetilde{U}_j)_{,j} + 2 \left \{\overline{U_i \times [p_{,i} - (\sigma_{ij})_{,j}]} - \widetilde{U}_i \times \overline{p_{,i} - (\sigma_{ij})_{,j}} \right \}  \\
    = -2 \tau_{ij} \widetilde{S}_{ij} - [\overline{\rho}(\widetilde{U_i U_i U_j} - \widetilde{U_i U_i} \widetilde{U}_j)]_{,j} + 2 (\tau_{ij} \widetilde{U}_i)_{,j}.
\end{aligned}
\end{equation}
In Eq.(\ref{substraction_ns_3}), substituting $\overline{p} = \overline{\rho} R \tilde{T}$ into pressure-gradient velocity correlation, leads to the following form
\begin{equation*} \label{pressure_gradient}
\begin{aligned}
    L_4  = &\overline{U_i \times p_{,i}} - \widetilde{U}_i \times \overline{p_{,i}}
         = [\overline{\rho} R (\widetilde{T U_i} - \widetilde{T} \widetilde{U}_i]_{,i} -(\overline{p U_{i,i}} - \overline{p} \widetilde{U}_{i,i}).
\end{aligned}
\end{equation*}
The term $L_5$ can be designed and decomposed as follows
\begin{equation*} \label{disipation}
\begin{aligned}
L_5  = &\overline{U_i \times (\sigma_{ij})_{,j}} - \widetilde{U}_i \times \overline{(\sigma_{ij})_{,i}} \\
     = & (\overline{\sigma_{ij} U_i} - \overline{\sigma}_{ij} \widetilde{U}_i)_{,j} - (\overline{\sigma_{ij} U_{i,j}} - \overline{\sigma}_{ij} \widetilde{U}_{i,j}),
\end{aligned}
\end{equation*}
Plugging $L_4$ and $L_5$ into Eq.(\ref{substraction_ns_3}), the SGS kinetic energy equation reads
\begin{equation}
\begin{aligned} \label{substraction_ns_4}
    (\overline{\rho} K_{sgs})_{,t}  &+ (\overline{\rho} K_{sgs} \widetilde{U}_j)_{,j} = - \tau_{ij} \widetilde{S}_{ij} \color{black} - (\overline{\sigma_{ij} U_{i,j}} - \overline{\sigma}_{ij} \widetilde{U}_{i,j}) + (\overline{p U_{k,k}} - \overline{p} \widetilde{U}_{k,k}) \\
    &+ [- \frac{1}{2}\overline{\rho} (\widetilde{U_i U_i U_j} - \widetilde{U_i U_i} \widetilde{U}_j) + \tau_{ij} \widetilde{U}_i + (\overline{\sigma_{ij} U_i} - \overline{\sigma}_{ij} \widetilde{U}_i) - \overline{\rho} R (\widetilde{T U_j} - \widetilde{T} \widetilde{U}_j)]_{,j}.
\end{aligned}
\end{equation}

In practice, two assumptions are introduced to decompose the total SGS dissipation rate into SGS solenoidal part and SGS dilational one. Firstly, assume that the kinematic viscosity $\nu$ is spatially uniform over the filter width, so that $\overline{\mu \phi} = \overline{\rho} \nu \tilde{\phi} = \overline{\mu} \tilde{\phi}$. In addition, for compressible turbulence, the assumption $\overline{\sigma}_{ij} = 2 \overline{\mu} (\widetilde{S}_{ij} - \delta_{ij} \widetilde{S}_{kk}/3)$ is adopted in previous literature \cite{vreman1995priori, martin2000subgrid}. Then, the total SGS dissipation rate $\varepsilon^{sgs}$ in Eq.(\ref{substraction_ns_4}) can be rewritten as
\begin{equation*}\label{dissipation_decomposition}
\begin{aligned}
    \varepsilon^{sgs} &= \overline{\sigma_{ij} U_{i,j}} - \overline{\sigma}_{ij} \widetilde{U}_{i,j} \\
                &= 2 \overline{\mu}(\widetilde{S_{ij} U_{i,j}} - \widetilde{S}_{ij} \widetilde{U}_{i,j}) - 2 \overline{\mu} (\widetilde{U_{k,k}^2} - \widetilde{U}_{k,k}^2)/3.
\end{aligned}
\end{equation*}
Using the fact $S_{ij} S_{ij} = \omega_i \omega_i/2 + U_{i,j} U_{j,i}$, the total dissipation rate $\varepsilon^{sgs}$ could be decomposed into SGS solenoidal dissipation rate $\varepsilon_s^{sgs}$ and SGS dilational dissipation rate $\varepsilon_d^{sgs}$ as follow
\begin{equation}\label{dissipation_decomposition_2}
\begin{aligned}
    \varepsilon_s^{sgs} &=  \overline{\mu} (\widetilde{\omega_i \omega_i} - \widetilde{\omega}_i \widetilde{\omega}_i), \\
    \varepsilon_d^{sgs} &=  2 \overline{\mu} (\widetilde{U_{i,j} U_{j,i}} - \widetilde{U}_{i,j} \widetilde{U}_{j,i}) - 2 \overline{\mu} (\widetilde{U_{k,k}^2} - \widetilde{U}_{k,k}^2)/3,\color{black} \\
\end{aligned}
\end{equation}
where $\omega_i =  \epsilon_{ijk} U_{k,j}$ is the vorticity and $\widetilde{\omega}_i =  \epsilon_{ijk} \widetilde{U}_{k,j}$, with the alternating tensor $\epsilon_{ijk}$. With the reasonable assumption $U_{i,j} U_{j,i} \approx U_{k,k}^2$ \cite{wilcox1998turbulence} (exactly in homogeneous turbulence), the SGS dilational dissipation rate $\varepsilon_d^{sgs}$ in Eq.(\ref{dissipation_decomposition_2}) can be approximated as
\begin{equation}\label{dissipation_decomposition_3}
\begin{aligned}
\varepsilon_d^{sgs} = 4 \overline{\mu} (\widetilde{U_{k,k}^2} - \widetilde{U}_{k,k}^2)/3.
\end{aligned}
\end{equation}
The difference between current derivation on dissipation rate as Eq.(\ref{dissipation_decomposition_2}) and Eq.(\ref{dissipation_decomposition_3}) and the reference literature \cite{chai2012dynamic} as Eq.(3.4), Eq.(3.5) and Eq.(3.8) should be pointed out. In the reference \cite{chai2012dynamic}, Eq.(3.4) represents the total dissipation rate instead of the solenoidal dissipation rate.

\begin{figure}[!h]
	\centering
	\includegraphics[width=0.45\textwidth]{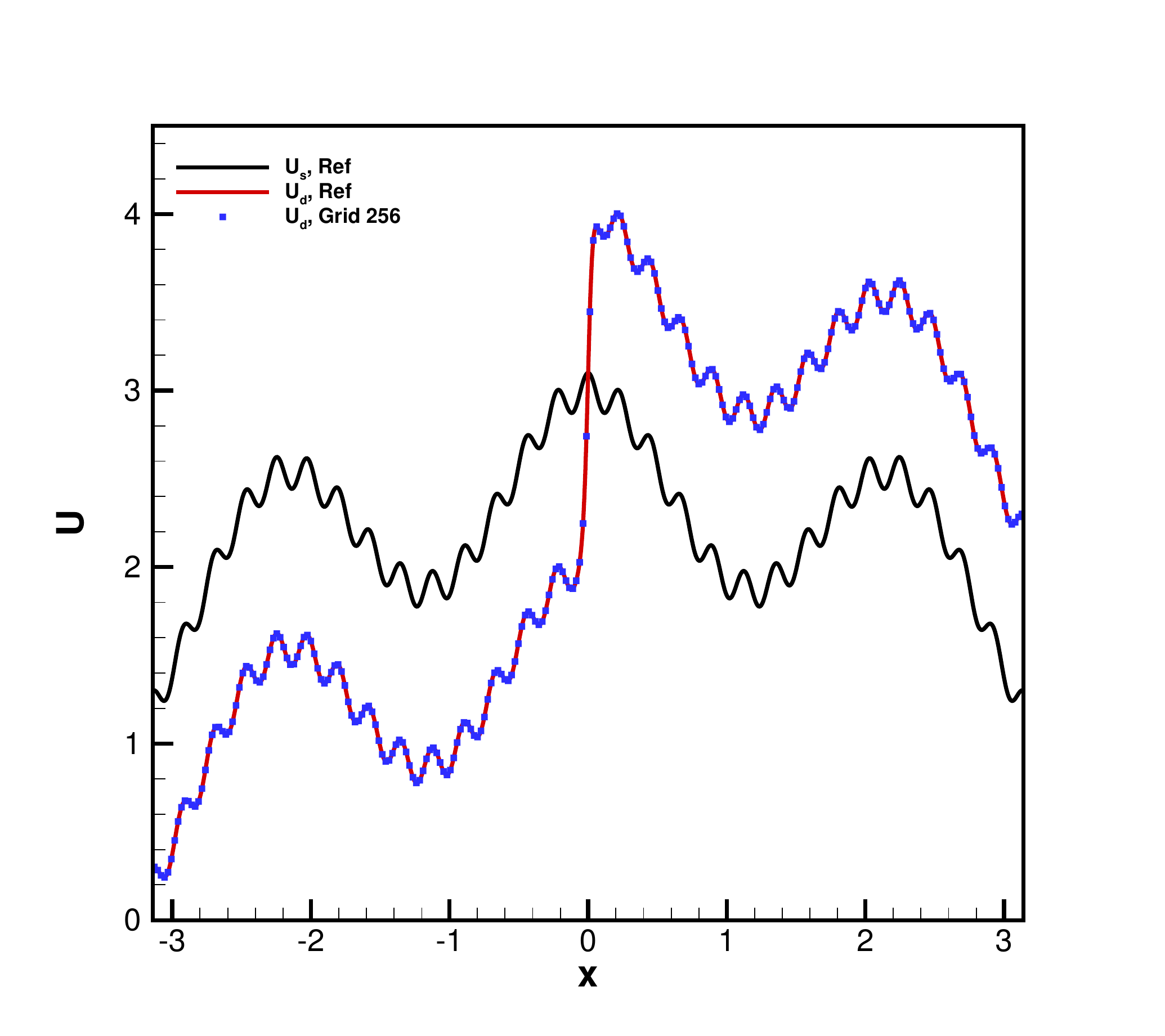}
	\includegraphics[width=0.45\textwidth]{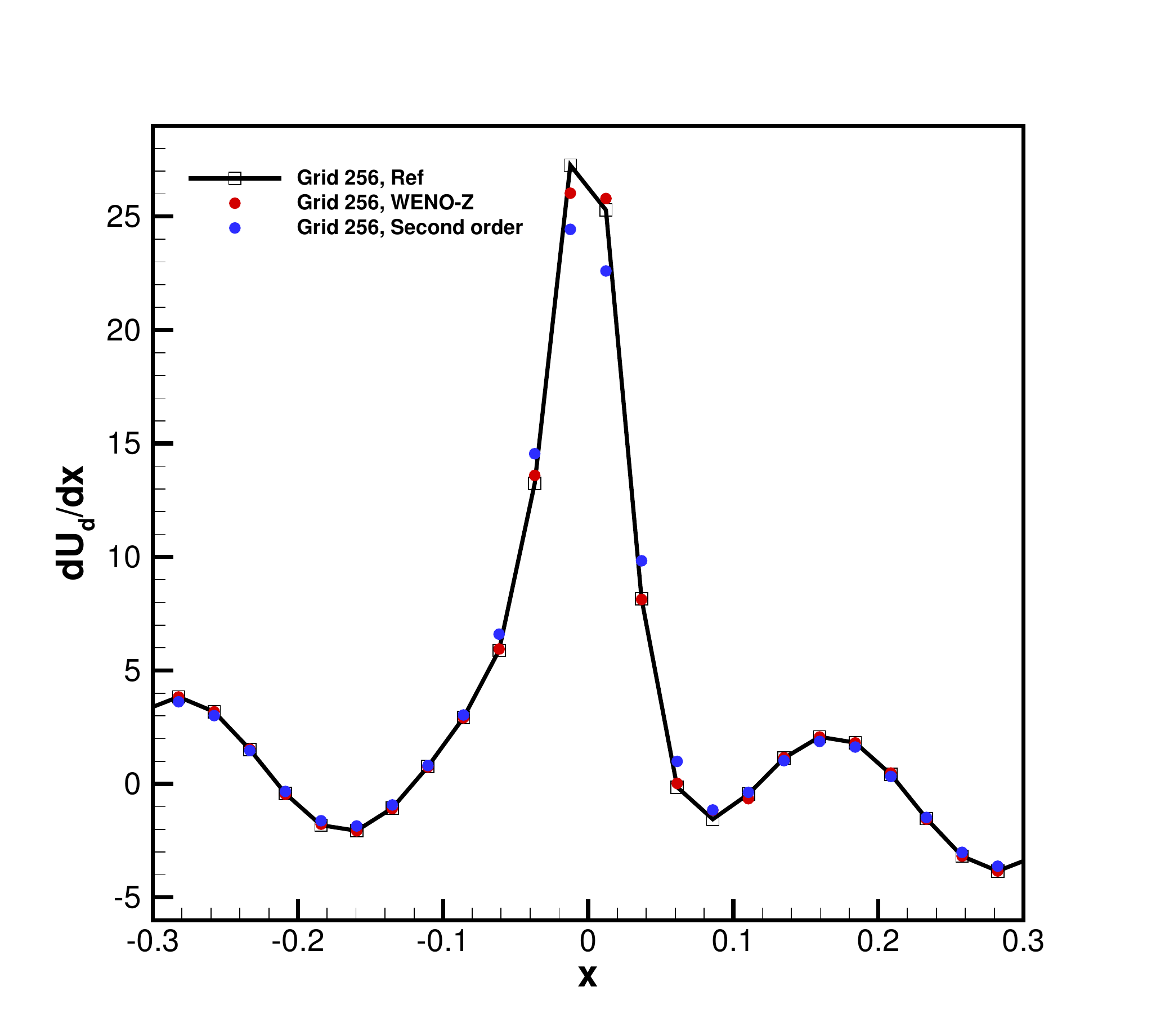}
	\vspace{-2mm}
	\caption{\label{figure_dU_ds} Initial wave of  $U_s(x)$ and $U_d(x)$, and spatial derivative of $U_d$(x) with the analytical solution, fifth-order WENO-Z reconstruction, and second-order central difference method.}
\end{figure}
\begin{table}[!h]
	\begin{center}
		\caption{\label{table_accuracy_test} Accuracy test of spatial derivative for $U_s(x)$ and $U_d(x)$ with WENO-Z reconstruction.}
		\vspace{3mm}
		\centering
		\begin{tabular}{c|cc|cc}
			\hline \hline
			Waves         &$U_s(x)$             &         &$U_d(x)$             & \\
			\hline
			Mesh length   &$L^2$ error          &Order    &$L^2$ error   &Order  \\
			\hline
			$2 \pi/64$    &$2.619209e-01$       &         &$3.458353e-01$       &       \\
			$2 \pi/128$   &$3.050078e-02$       &$3.10$   &$4.959606e-02$       &$2.81$ \\
			$2 \pi/256$   &$3.088122e-03$       &$4.62$   &$6.520176e-03$       &$4.14$ \\
			$2 \pi/512$   &$6.699721e-05$       &$3.98$   &$4.571909e-04$       &$3.19$ \\
			\hline \hline
		\end{tabular}
	\end{center}
	\vspace{-2mm}
\end{table}

\section*{Appendix B: spatial derivatives in consistent with numerical scheme}
The one-dimensional multiple-frequency smooth wave $U_s(x)$ as well as the waves with sharp derivative $U_d(x)$ are used to test the accuracy of spatial derivative.
The sharp derivative is designed for simulating the shocklets as shown in Fig.{\ref{q_pdftheta_R1R2}}. The $U_s(x)$ and $U_d(x)$ are given by
\begin{equation}\label{U_sdx}
\begin{aligned}
U_s(x) &= \sum_{i = 1}^{3} \alpha_i cos(2\beta_i\pi x), x \in [-\pi, \pi],\\
U_d(x) &= \sum_{i = 1}^{3} \alpha_i cos(2\beta_i\pi x) + \tanh(\gamma x), x \in [-\pi, \pi],\\
\end{aligned}
\end{equation}
where the coefficients $\alpha_1 = 800, \alpha_2 = 80, \alpha_1 = 8$ and $\beta_1 = 0.1, \beta_1 = 0.5, \beta_1 = 2.5, \gamma = 30$ are adopted.
Initial waves of $U_s(x)$ and $U_d(x)$ are presented in Figure \ref{figure_dU_ds}.
Three method are used to compute the spatial derivative, namely the analytical solution, fifth-order WENO-Z reconstruction \cite{castro2011high}, and second-order central difference method.
Compared with the analytic solution, the fifth-order WENO-Z reconstruction outweighs the second-order central difference method.
In current paper, WENO-Z reconstruction is applied to compute the spatial derivative.
The accuracy tests of $U_s(x)$ and $U_d(x)$  with WENO-Z reconstruction are shown in Table.\ref{table_accuracy_test}.
Here, the WENO-Z reconstruction for spatial derivatives is consistent with the HGKS when obtaining the high-fidelity DNS data \cite{cao2019three}.

\bibliographystyle{unsrt}
\bibliography{caogybib}
\end{document}